\documentclass[preprint,letterpaper,prc,aps,showpacs,showkeys]{revtex4}

\usepackage{graphicx}
\usepackage{tabularx}
\newcommand{\beq}{\begin{equation}}
\newcommand{\eeq}{\end{equation}}
\newcommand{\beqa}{\begin{eqnarray}}
\newcommand{\eeqa}{\end{eqnarray}}
\newcommand{\nn}{\nonumber}
\newcommand{\Sigs}{\Sigma_{\mathrm s} }
\newcommand{\Sigv}{\Sigma_{\mathrm v} }
\newcommand{\Sigo}{\Sigma_{\mathrm o} }

\newcommand{\kf}{k_{\mathrm F} }

\newcommand{\bfgamma}{\mbox{\boldmath$\gamma$\unboldmath}}

\newcommand{\veck}{\textbf{k}}

\newcommand{\vecq}{\textbf{q}}

\newcommand{\qabs}{|{\bf q}|}
\newcommand{\kabs}{|{\bf k}|}
\makeatletter
\newcommand{\fmslash}[2][0mu]{%
  \mathchoice
    {\fmsl@sh\displaystyle{#1}{#2}}%
    {\fmsl@sh\textstyle{#1}{#2}}%
    {\fmsl@sh\scriptstyle{#1}{#2}}%
    {\fmsl@sh\scriptscriptstyle{#1}{#2}}}
\newcommand{\fmsl@sh}[3]{%
  \m@th\ooalign{$\hfil#1\mkern#2/\hfil$\crcr$#1#3$}}

\newcommand{\fet}[1]{\mbox{\boldmath $#1$}}

\makeatother

\begin{document}
\preprint{}
\title{Nuclear matter in the chiral limit and the in-medium chiral condensate}
\author{O. Plohl}
\author{C. Fuchs}
\affiliation{Institut
f$\ddot{\textrm{u}}$r Theoretische Physik, Universit$\ddot{\textrm{a}}$t
T$\ddot{\textrm{u}}$bingen,
Auf der Morgenstelle 14, D-72076 T$\ddot{\textrm{u}}$bingen, Germany}
\begin{abstract}
 We investigate nuclear matter, i.e. the nuclear equation-of-state (EOS) 
as well as the relativistic mean fields in the chiral limit. The investigations are 
based on a chiral nucleon-nucleon EFT interaction where the explicit and implicit pion mass 
dependence is known up to next-to-leading order. The nuclear bulk 
properties are found to remain fairly stable in the chiral limit. 
Based on the same interaction the 
in-medium scalar condensate is derived, both in Hartree-Fock approximation as well 
as from the Brueckner G-matrix, making thereby use of the Hellman-Feynman 
theorem. Short distance physics which determines the reduction of the 
in-medium nucleon mass is found to play only a minor role for the 
reduction of the chiral condensate.

\end{abstract}
\maketitle

%%%%%%%%%%%%%%%%%%%%%%%%%%%%%%%%%%%%%%%%%%%
\section{Introduction}
%%%%%%%%%%%%%%%%%%%%%%%%%%%%%%%%%%%%%%%%%%%

The investigation of dense nuclear matter is one of the most exciting topics 
of present days nuclear and hadron physics. Since hadrons are excitations 
of the QCD vacuum their in-medium excitation spectrum  is closely connected 
to the modification of the QCD vacuum at finite density. The chiral 
condensate  $\langle {\bar q}q \rangle$ is expected to be reduced at 
finite density and/or temperature which 
leads to a partial restoration of chiral symmetry of QCD and should 
be reflected, e.g., in shifts of the corresponding hadron masses. 
Chiral symmetry is an exact symmetry of QCD 
in the limit of massless quarks.  
Since the up and down current quark masses 
are small, i.e. of the order of 5-10 MeV, 
this symmetry is still approximately fulfilled. In nature it is, 
however, spontaneously broken by the 
non-vanishing -- and large -- expectation value of the scalar quark 
condensate  $\langle {\bar q}q \rangle$ of the QCD vacuum. 
The spontaneous symmetry breaking, similar to the 
spontaneous magnetization of a ferromagnet which breaks the symmetry of the 
underlying Hamiltonian, implies the existence of massless Goldstone 
bosons which are the pions. The small pion mass of 140 MeV ensures that 
the concept of chiral symmetry persists as a fundamental feature of low 
energy hadron physics. 

To leading order in density the in-medium scalar condensate drops linearly 
with the nuclear density with the proportionality given by the pion-nucleon 
sigma term $\sigma_N$. The model independent low density behavior can 
be derived from the Hellmann-Feynman theorem \cite{cohen91}. To determine the 
corrections to this leading density dependence a large variety of models 
has been exploited. These were e.g. the Nambu-Jona-Lasinio (NJL) model 
\cite{cohen91,weise90,tuebingenC}, various versions of the linear 
sigma model \cite{Chanfray05,chanfray06}, the Quark Meson Coupling model (QMC) 
\cite{Saito05} or recently the Polyakov-NJL model \cite{Ratti05}. The 
Hellmann-Feynman theorem relates the in-medium  scalar quark 
condensate with the quark mass derivative of the total energy density. 
The latter quantity can also be calculated within hadron effective field 
theory \cite{cohen91,brockmann96,Li94}, such as $\sigma\omega$ type models \cite{sw86} and, more
microscopically, within nuclear many-body theory. In a relativistic framework
the  latter leads to the relativistic Brueckner (DBHF) approach
\cite{terhaar87a,bm90,gross99} where the nucleon-nucleon interaction is 
given by boson-exchange models \cite{bonn}. While DBHF allows a quite 
reliable determination of the nuclear equation of state (EOS) up to 
at least two times nuclear density as a conservative estimate - 
recent DBHF calculations \cite{dalen04} have been shown to be consistent with 
astrophysical accelerator based constraints concerning their high density 
behavior \cite{klaehn06} - the unknown quark mass dependence of the mesonic
couplings and masses introduces large errors when conclusions on the 
in-medium condensate are drawn \cite{brockmann96,Li94}.

Most hadronic models do, however,  not respect chiral symmetry. 
A more systematic and direct connection to QCD is provided by 
chiral effective field theory (EFT). Up to now the two-nucleon system 
has been considered up to next-to-next-to-next-to-leading order (N$^3$LO) in 
chiral perturbation theory \cite{entem02,entem03,epelbaum05}. 
In such approaches the nucleon-nucleon ($NN$) potential consists 
of one-, two- and three-pion exchanges and 
contact interactions which account for the short-range contributions. 
The advantage of such approaches is the systematic expansion of the $NN$ 
interaction in terms of chiral power counting. The expansion is performed 
in powers of $(Q/\Lambda_\chi)^\nu$ where $Q$ is the generic low 
momentum scale given by the nucleon three-momentum, or the four-momenta of 
virtual pions or a pion mass. 
$\Lambda_\chi \simeq 4\pi f_\pi \simeq 1$ GeV is the chiral
symmetry breaking scale.  Moreover, chiral EFT has a well 
defined quark mass dependence. For the $NN$ interaction the chiral limit 
of vanishing current quark and pion masses has been evaluated up 
to next-to-leading order (NLO)~\cite{epelbaum03,Beane:2002xf} and applied 
to the deuteron problem \cite{epelbaum03b}.

Exploiting the Hellmann-Feynman theorem chiral EFT allows thus a precise 
determination of the scalar condensate at NLO. 

Within finite density QCD sum rules the  scalar condensate determines 
automatically the shift of the effective nucleon mass $M^*$. 
In the sum rule approach scalar and vector fields arise 
naturally from the structure of the quark propagator which is 
proportional to the corresponding condensates. The quark correlation function 
can be expressed to leading order in terms of the scalar condensate 
$\langle {\bar q}q  \rangle_{\rho_B} $ 
and the vector condensate $\langle q^\dagger q  \rangle_{\rho_B} $ 
which is introduced by the breaking of Lorentz invariance due to the presence 
of the medium. The identification of the correlation function with the 
in-medium nucleon propagator of a dressed quasi-particle leads 
to scalar and vector self-energies 
$\Sigs$ and $\Sigo$ which are of the same order in the 
condensates~\cite{cohen91,drukarev91}.

The scalar and vector self-energies 
$\Sigs$ and $\Sigo$ are on the other hand key quantities of each 
relativistic hadronic field theory. They determine the nuclear EOS, 
the effective nucleon mass $M^*=M+\Sigs$, the single particle potential 
$U_{\rm cent} \simeq \Sigo + \Sigs \simeq -50$ MeV and the 
spin-orbit potential $U_{\rm S.O.} \propto  (\Sigo -  \Sigs) \simeq 750$ MeV 
in finite nuclei.  Relativistic phenomenology implies large fields 
of opposite sign $\Sigs\simeq -350$ MeV and 
$\Sigo \simeq +300$ MeV \cite{sw86,rmf}. The same is obtained 
by relativistic many-body theory, 
i.e. DBHF \cite{terhaar87a,bm90,gross99,dalen04}.  
One has, however,  to keep in mind that the individual scalar and 
vector components are interpolating fields which do not directly 
manifest in experimentally accesable observables. Nevertheless,   
when modern nucleon-nucleon interactions are mapped on a Lorentz 
covariant operator basis (using projection techniques as briefly 
described in Section~\ref{selfenergy}) they reveal large scalar and 
vector fields of comparable size in nuclear matter as a model 
independent fact \cite{plohl06,plohl06_2}.
This holds not only for manifestly covariantly formulated interactions, 
such as relativistic One-Boson-Exchange models (Bonn, CD-Bonn, Nijmegen), 
but also for non-relativistic interactions 
(Argonne $v_{18}$, Reid93, Idaho N$^3$LO, $V_{\rm low~k}$) 
as soon as the symmetries of the Lorentz 
group are restored. From the analysis of the chiral EFT (Idaho N$^3$LO) 
potential \cite{entem02,entem03} we found that these fields are generated  
mainly by contact terms which occur at next-to-leading order in the chiral 
expansion and which generate the short-range spin-orbit 
interaction \cite{plohl06_2}. 

At moderate 
nuclear densities the N$^3$LO scalar and vector fields 
were found to be in almost perfect agreement with the prediction 
from leading order QCD sum rules \cite{plohl06_2}. The coincidence 
of the nucleon mass shifts obtained from QCD sum rules and relativistic 
nuclear phenomenology has been stressed in many works. However, 
this agreement is not yet fully understood. As pointed out in 
~\cite{birse96} a naive direct dependence of the 
nucleon mass $M^*$ on the scalar quark condensate 
leads to contradictions with chiral 
power counting. Long-distance physics from virtual pions,~i.e., 
the non-analytic term in 
the expansion of $\sigma_N$ gives a sizable contribution to the modification 
of the in-medium quark condensate. Such contributions are, however, 
found to play only a minor role for the reduction of the nucleon mass. 

In the present work we calculate both quantities, the scalar quark 
condensate and the effective nucleon mass $M^*$ from the same chiral 
effective interaction. The condensate is determined making use of the 
  Hellmann-Feynman theorem and the fact that, at least up to NLO, the 
quark mass dependence of the potential is known from its analytic 
formulation in the chiral limit. The effective mass, on the other 
hand, can be determined in Hartree-Fock approximation making use 
of projection techniques on a relativistic operator basis 
\cite{plohl06,plohl06_2}. 

Since the quark mass dependence is known for the chiral effective interaction
up to NLO this allows furthermore to investigate possible changes of 
the properties of symmetric nuclear matter up to this order. 
Therefore the 
large attractive scalar and repulsive vector self-energy components 
as well as the nuclear equation of state are studied in the chiral limit.
This has also been done using projection techniques on a relativistic 
operator basis in Hartree-Fock approximation. Naturally one assumes that
the magnitude of these fields persists even in the chiral limit
since hadronic properties are not expected to change dramatically in the case
of massless quarks or pions. The reason for this is that the expansion of the
nuclear force in the context of chiral perturbation theory is well defined
for small quark masses and should still be valid in the limit 
$m_q \rightarrow 0$ which is equivalent to $m_\pi \rightarrow 0$.

In the first part of the paper a short description of the chiral effective
$NN$ interaction is presented.
This is
followed by the formalism for the relativistic self-energy components
in Hartree-Fock approximation.
The determination of the relativistic self-energy is based on a 
projection technique which allows to transform any two-body potential 
amplitude onto a covariant operator basis or, in other words, to restore the 
symmetries of the Lorentz group.
In the following section the implications for the self-energy components and 
the equation of state for nuclear matter are shown when going to the 
chiral limit. 
In the last section we first present the prediction of the scalar quark 
condensate 
in matter derived with the help of the Hellmann-Feynman theorem. This  
is done within two different approximations, namely Hartree-Fock and the 
Brueckner ladder approximation. The latter allows to study the influence of 
short range correlations. Finally the connection between the 
effective nucleon mass and the scalar quark condensate is discussed.

%%%%%%%%%%%%%%%%%%%%%%%%%%%%%%%%%%%%%%%%%%%

%%%%%%%%%%%%%%%%%%%%%%%%%%%%%%%%%%%%%%%%%%%

%%%%%%%%%%%%%%%%%%%%%%%%%%%%%%%%%%%%%%%%%%%
\section{The EFT nucleon-nucleon interaction}\label{chiralNLO}
%%%%%%%%%%%%%%%%%%%%%%%%%%%%%%%%%%%%%%%%%%% 
Both, the investigation of the structure of the self-energy in 
nuclear matter and the resulting EOS in the chiral limit ($m_\pi \rightarrow 0$) 
as well as the determination
of the scalar condensate in matter depend crucially on the exact 
knowledge of the
implicit and explicit current quark mass dependence of the nuclear force 
on which the many-body approaches are based on. We are now in the situation to make use of the 
chiral $NN$ interaction derived in~\cite{epelbaum03} which 
allows an extrapolation in the pion mass where the quark mass dependence 
is known analytically up to NLO .

The chiral EFT potential  consists of one-, two- and three-pion exchanges and 
regularizing contact interactions describing the short-range contributions. 
The chiral expansion of the $NN$ 
interaction is performed  by organizing the contributions in terms of 
powers of $(Q/\Lambda_\chi)^\nu$ where $Q$ is the generic low 
momentum scale given by the nucleon three-momentum, or the four-momenta of 
virtual pions or a pion mass and $\Lambda_\chi \simeq 4\pi f_\pi \simeq 1$ GeV 
is the chiral symmetry breaking scale. 

In~\cite{epelbaum03} the light quark mass dependence of the nuclear force 
 has been derived up to next-to-leading (NLO) order in the framework of a 
modified Weinberg power counting,~i.e., additionally to the one-pion exchange 
(OPE) potential and contact terms the leading two-pion exchange (TPE) has been 
considered.

The explicit form of the chiral
effective $NN$ potential $V_{\rm NLO}$ we use is given by~\cite{epelbaum03}
\beq
V_{\rm NLO} = V^{\rm OPE} + V^{\rm TPE} + V^{\rm cont}\,,
\eeq
where 
\beqa
\label{potfin}
V^{\rm OPE} &=& - \frac{1}{4} \frac{g_A^2}{f_\pi^2} \left( 1 + 2 \Delta
- \frac{4 \tilde m_\pi^2}{g_A} \bar d_{18} \right) 
\, \fet \tau_1 \cdot \fet \tau_2 \, \frac{(\vec \sigma_1 \cdot \vec q \,) 
( \vec \sigma_2 \cdot \vec q\,)}
{\vec q\, ^2 + \tilde m_\pi^2} ~,\\ 
\label{potfinTPE}
V^{\rm TPE} &=& - \frac{ \fet{\tau}_1 \cdot \fet{\tau}_2 }{384 \pi^2 
f_\pi^4}\,\biggl\{
L(q) \, \biggl[4\tilde m_\pi^2 (5g_A^4 - 4g_A^2 -1) + \vec q\, ^2 (23g_A^4 - 
10g_A^2 -1) 
+ \frac{48 g_A^4 \tilde m_\pi^4}{4 \tilde m_\pi^2 + \vec q\, ^2} \biggr] \nn\\
&& {} \mbox{\hskip 2 true cm}+ \vec q\, ^2 \, \ln \frac{\tilde m_\pi}{m_\pi} 
\, (23g_A^4 - 10g_A^2 -1) \biggr\} \\
&& - \frac{3 g_A^4}{64 \pi^2 f_\pi^4} \,\left( L(q) + 
\ln \frac{\tilde m_\pi}{m_\pi} \right)\, \biggl\{
\vec{\sigma}_1 \cdot\vec{q}\,\vec{\sigma}_2\cdot\vec{q} - \vec q\,^2 \, 
\vec{\sigma}_1 \cdot\vec{\sigma}_2 \biggr\}~, \nn\\
\label{potfincont}
V^{\rm cont} &=& \bar C_S + \bar C_T (\vec \sigma_1 \cdot \vec \sigma_2 ) 
+ \tilde m_\pi^2 \, \left( \bar D_S - \frac{3 g_A^2}{32 \pi^2 f_\pi^4} (8 f_\pi^2  C_T- 5 g_A^2 + 2) 
\ln \frac{\tilde m_\pi}{m_\pi} \right) \\
&& {}+ \tilde m_\pi^2 \left( \bar D_T - \frac{3 g_A^2}{64 \pi^2 f_\pi^4} ( 16 f_\pi^2  C_T- 5 g_A^2 + 2) 
\ln \frac{\tilde m_\pi}{m_\pi} \right)\,
(\vec \sigma_1 \cdot \vec \sigma_2 ) \nn\\
&& {} + C_1 {\vec q \,}^2 
+ C_2 {\vec k \,}^2 + ( C_3 {\vec q \, }^2 + C_4 {\vec k \,}^2 ) 
( \vec \sigma_1 \cdot \vec \sigma_2 ) \nn\\
&& {} + i C_5 \frac{ \vec \sigma_1 + \vec \sigma_2}{2} 
\cdot ( \vec k \times \vec q \, ) 
+ C_6 ( \vec q \cdot \vec \sigma_1 ) 
( \vec q \cdot \vec \sigma_2 ) + C_7 ( \vec k \cdot \vec \sigma_1 )
( \vec k \cdot \vec \sigma_2 ) \,,
\nonumber
\eeqa
with $g_A$ and $f_\pi$ the physical values of the nucleon axial coupling and 
the pion decay constant, respectively. Because at NLO any 
shift in $g_A$ and $f_\pi$ for a different value of $m_\pi$ in the TPE 
is a N$^4$LO effect for the TPE the physical values $g_A=1.26$ and 
$f_\pi=92.4$ MeV are used. 
The value of the pion mass is indicated by $\tilde m_\pi$ 
compared to the physical one denoted by $m_\pi$. $L(q)$ is given by 
\beq
L(q) \equiv L(| \vec q \,|) 
= \frac{\sqrt{4 \tilde m_\pi^2 + \vec q\, ^2}}{|\vec q \,|} \, 
\ln \frac{ \sqrt{4 \tilde m_\pi^2 + \vec q\, ^2}
+ | \vec q \, |}{2 \tilde m_\pi}\, .
\eeq
$\Delta$ represents the relative shift in the ratio $g_A/f_\pi$ compared to its 
physical value since they show an implicit dependence on the pion mass
\beqa
\label{deltaCL}
\Delta &\equiv& \frac{\left(g_A/f_\pi\right)_{\tilde m_\pi}-\left(g_A/f_\pi\right)_{m_\pi}}{
\left(g_A/f_\pi\right)_{m_\pi}} \\
&=& \left( \frac{g_A^2 }{16 \pi^2 f_\pi^2} - \frac{4 }{g_A}
\bar{d}_{16} + \frac{1}{16 \pi^2 f_\pi^2} \bar{l}_4 \right) (m_\pi^2 - \tilde m_\pi^2) - 
\frac{g_A^2 \tilde m_\pi^2}{4 \pi^2 f_\pi^2} \ln \frac{\tilde m_\pi}{m_\pi} \,.
\nonumber
\eeqa

The low-energy constants (LECs) $\bar C_{S,T}$ and $\bar D_{S,T}$ appear at LO and 
are related 
to the $C_{S,T}$ from \cite{epelbaum05} via
\beq
C_{S,T} = \bar C_{S,T} + m_\pi^2 \bar D_{S,T}\,.
\eeq

The LECs $\bar D_{S,T}$ have not been fixed by experiment till now. 
In Ref.~\cite{epelbaum03} natural values have been assumed for these constants
\beq
\label{DbarST}
\bar D_{S,T}=\frac{\alpha_{S,T}}{f^2_\pi\Lambda_\chi^2}\, ,
\quad \mbox{where} \quad \alpha_{S,T} \sim 1 \quad \mbox{and}
\quad \Lambda_\chi \simeq 1\,{\rm GeV}.
\eeq
The LECs $\bar d_{16}$,$\bar d_{18}$ and $\bar l_4$ are related to 
pion-nucleon interactions. We take $\bar l_4=4.3$ which is fixed 
with relatively small error bars.
The LECs $\bar d_{16}$,$\bar d_{18}$ are not yet uniquely fixed, i.e. there
exists a certain range of possible values fixed from different observables.
The implications on the results induced by the uncertainties of the LECs 
will be discussed later in greater detail.

%%%%%%%%%%%%%%%%%%%%%%%%%%%%%%%%%%%%%%%%%%%
\section{Covariant representation and the nucleon self-energy}\label{selfenergy}
%%%%%%%%%%%%%%%%%%%%%%%%%%%%%%%%%%%%%%%%%%%

In the following the mean field in nuclear matter is determined 
 by calculating the relativistic self-energy $\Sigma$ in Hartree-Fock 
approximation at {\it tree level}. It provides therefore a qualitative rather
than a quantitative description of the nuclear many-body problem. 
Nevertheless the scale of these fields is already set at tree 
level. Although 
essential for nuclear binding and saturation, higher order correlations, 
in particular short-range correlations, change the size of the fields by 
less than 25\%, as has been estimated in ~\cite{plohl06_2} comparing 
HF to a full self-consistent relativistic DBHF calculation. 
However, for nuclear binding 
these deviations are essential. To meet 
the empirical saturation point of nuclear matter one has to
introduce a self-consistent scheme and 
to account for short-range correlations.  Such calculations have to 
be based on the in-medium T-matrix (or G-matrix) rather than the 
bare potential $V$. 
G-matrix correlations will be discussed in connection 
with the chiral condensate in the next section. 

Here we shortly sketch the formalism for the evaluation of the self-energy. 
More details can be found in Refs.~\cite{plohl06_2} and~\cite{gross99}.

The self-energy for a nucleon with four-momentum $k$ follows from the 
interaction $V$ by integrating over the occupied states $q$ in the Fermi sea
\beq
\Sigma_{\alpha\beta}(k,k_F)=-i\int {d^4q\over {(2\pi)^4}}~
G^D_{\tau\sigma}(q)~V^A(k,q)_{\alpha\sigma;\beta\tau}~.
\label{sig:hf}
\eeq
The evaluation of the Hartree integral is sufficient because 
the matrix elements are fully anti-symmetrized containing 
direct (Hartree) and exchange (Fock) contributions.
To determine the self-energy only positive-energy states are taken into 
account as done in the standard DBHF approach.
The Dirac propagator 
\beq
G^D(q)=[\fmslash{q}+M]2\pi
 i\delta(q^2-M^2)\Theta(q_0)\Theta(k_F-\qabs)
\eeq
describes the on-shell propagation
of a nucleon with momentum ${\bf q}$ and energy $E_{\bf q}=\sqrt{{\bf
 q}^2+M^2}$ inside the Fermi sea. Due to the $\Theta$ functions in the 
propagator only positive energy nucleons are allowed in the intermediate 
scattering states which prevents the occurence of divergent contributions 
coming from negative energy states.

Based on symmetry considerations in isospin saturated symmetric nuclear matter
the self-energy can be written as a sum  of a scalar $\Sigs$, a time-like 
vector $\Sigo$ and a spatial vector part $\Sigv$.
Thus, in nuclear matter rest frame
the Dirac structure of the self-energy has the simple form
\beqa
\Sigma(k,\kf)= \Sigs (k,\kf) -\gamma_0 \, \Sigo (k,\kf) + 
\bfgamma  \cdot \textbf{k} \,\Sigv (k,\kf).
\label{subsec:SM;eq:self1}
\eeqa

The Lorentz components of the self-energy operator (\ref{sig:hf}) are then 
given by~\cite{gross99}
\beqa
\Sigs  & = & \frac{1}{4} \int^{k_F} \,\frac{d^3\vecq}{(2
  \pi)^3}  \frac{M}{E_{\bf q}}  \left[ 4g_{\rm S} -
  g_{\tilde {\rm S}}+ 4 g_{\rm A} -\frac {(k^{\mu}-q^{\mu})^2} {4M^2}
  g_{\widetilde {\rm PV}} \right]~, \nonumber\\
\Sigo   & = &  \frac{1}{4} \int^{k_F}\, \frac{d^3\vecq}{(2 \pi)^3}    \left[ g_{\tilde {\rm S}}- 2 g_{\rm A}+ \frac{E_{\bf k}}{E_{\bf q}} \frac {(k^{\mu}-q^{\mu})^2} {4M^2} g_{\widetilde {\rm PV}}\right]
\label{sigHF}\\
\Sigv  & = & \frac{1}{4} \int^{k_F}\, \frac{d^3\vecq}{(2
  \pi)^3} \frac{\veck\cdot\vecq}{\kabs^2E_{\bf    q}}  
\left[ g_{\tilde {\rm S}}-2 g_{\rm A} + \frac{k_z}{q_z} \frac{(k^{\mu}-q^{\mu})^2} {4M^2}
  g_{\widetilde {\rm PV}} \right]
\nonumber 
\eeqa

To evaluate the self-energy operator in the nuclear matter rest frame
the two-body interaction matrix 
$V$ determined in the two-particle centre-of-mass (c.m.) 
frame and usually given in 
the $|JLS\rangle$ basis, has to be represented covariantly by 
Dirac operators and Lorentz invariant amplitudes~\cite{tjon85}, a procedure
 which can be applied to any two-body amplitude.
This is also the most convenient way to Lorentz-transform the interaction 
matrix from one frame into another~\cite{horowitz87}.
 
Naturally a fully relativistic treatment invokes 
the excitation of anti-nucleons. However, standard $NN$ potentials 
(even OBE type 
potentials such Bonn, CD-Bonn or Nijmegen) are restricted to the positive 
energy sector and neglect the explicit coupling to anti-nucleons. 
As a consequence 
one has to work in a subspace of the full Dirac space. This shortcoming can 
be avoided using fully covariant potentials which explicitely include 
anti-nucleon states \cite{gross,Gross:2007be}. The present investigations 
and those in Refs. \cite{plohl06,plohl06_2} have, however, been restricted 
to "standard" potentials based on the no sea approximation. Similarly, 
the EFT potentials \cite{entem02,entem03} applied here and in our previous 
investigations do not explicitly include anti-nucleons, in 
contrast to covariant approaches which require renormalization 
procedures to restore
chiral power counting \cite{Lehnhart:2004vi}.

Working in the positive energy subspace, symmetry arguments
and the restriction to on-shell scattering allow the two-body matrix elements
to be are expanded in terms of five Lorentz invariants.
A possible choice of a set of five linearly independent operators are
the scalar, vector, tensor, axial-vector and 
pseudo-scalar Fermi covariants $\Gamma_m =\{{\rm S,V,T,P,A}\}$ with 
\beq
 {\rm S}=1\otimes1,\quad {\rm V}=\gamma^{\mu}\otimes\gamma_{\mu},\quad
  {\rm T}=\sigma^{\mu\nu}\otimes \sigma_{\mu\nu},\quad
{\rm P}=\gamma_5\otimes\gamma_5,\quad {\rm A}=\gamma_5\gamma^{\mu}\otimes \gamma_5\gamma_{\mu}. 
\label{cov1}
\eeq
 
The choice of the operator basis is not unique. 
In~\cite{gross99} it has been shown that the so-called complete 
$pv$ representation is an appropriate choice
where the set of covariants originally proposed 
by Tjon and Wallace \cite{tjon85} is given by 
\beq
\Gamma_m = \{ {\rm S},-{\rm {\tilde S}},({\rm A}-  {\rm {\tilde A}}),{\rm PV} ,
-{\rm {\widetilde {PV}}}\}~.
\label{cov3}
\eeq
PV and ${\rm {\widetilde {PV}}}$ are the direct and exchange 
pseudo-vector covariants, analogous to the pseudo-scalar covariant P, 
however, with $\gamma_5$ replaced by 
$(\fmslash{q}^\prime - \fmslash{q})/2M \gamma_5$. A PV vertex suppresses the 
coupling to negative states and it is consistent with soft pion
theorems based on chiral symmetry considerations.

Thus the on-shell 
($|{\bf q}|=|{\bf q}^\prime|$) scattering matrix is given by 
\begin{equation}
{\hat V}^I ({\bf q}^\prime ,{\bf q})
=\sum_{m} g_{m}^I(|{\bf q}|,\theta)~ \Gamma_m~~,
\label{vcov}
\end{equation}
where $\theta$ is the c.m. scattering angle and $I=0,1$ the 
isospin channel. For the Hartree-Fock self-energy 
it is sufficient to consider $\theta=0$  
when anti-symmetrized matrix elements are used since 
 $\theta=\pi$ contains then only redundant information. 
The transformation of the Born 
amplitudes from an angular-momentum basis onto 
the covariant basis (\ref{vcov}) is now standard and 
runs over the following steps:  
$ |LSJ\rangle \rightarrow$ {\it partial wave helicity states 
$ \rightarrow$ plane wave helicity states $\rightarrow$ 
covariant basis}. The first two transformation can e.g. be found in Refs. 
\cite{machleidt89}. The last step has to be performed numerically 
by matrix inversion  \cite{horowitz87,gross99}.  

The expressions for the NLO chiral potential given in  
Eqs.~(\ref{potfin})-(\ref{potfincont})
are independent of the nucleon mass. Nevertheless the nucleon mass appears 
in the expressions for the calculation of the relativistic 
mean fields, Eq.~(\ref{sigHF}), as well as in the procedure of projecting 
the two-body amplitudes on the covariant basis. Moreover, the quark mass 
dependence of the nucleon mass is also the leading term which determines 
the quark mass dependence of the EOS. The EOS, i.e. the 
energy per particle E/A is given by 
\beq
 {\rm E / A }=\frac{1}{\rho_B} \int_F \frac{d^3{\bf k}}{2\pi^3} 
    \left[\frac {k^2}{2M}+\frac{1}{2}\,U_{\rm s.p.} (k,\kf) \right]  
\label{eos1}
\eeq
with the single particle potential $U_{\rm s.p.} (k,\kf)$  
defined through the fields
\begin{eqnarray}
U_{\rm s.p.} (k,\kf) = \frac{M}{E} \Sigs - \frac{ k_{\mu} \Sigma^\mu}{E} 
         = \frac{M \Sigs }{\sqrt{ {\bf k}^2 + M^{2}}} 
         - \Sigo + \frac{ \Sigv {\bf k}^2}
           {\sqrt{ {\bf k}^2  + M^{2}}}
\quad .
\label{upot2}
\end{eqnarray}
The integration in Eq. (\ref{eos1}) runs over the Fermi sea $F$ and 
we account for the full momentum dependence of the self-energy 
components $\Sigs,\Sigo,\Sigv$. 

Within the framework of chiral EFT the physical (vacuum) 
nucleon mass $M$ can be expressed as
\beq
M=M_0+\sigma_N
\label{nmass1}
\eeq
where $M_0$ is the value of the nucleon mass in the chiral limit. 
The sigma term $\sigma_N$ represents the contribution from explicit chiral
symmetry breaking to the nucleon mass and determines the 
quark mass dependence of the nucleon mass
\beqa
\sigma_N = \sum _{q=u,d} m_q\frac{dM}{dm_q} =  m_\pi^2\frac{dM}{dm_\pi^2}  
\label{sigma1}
\eeqa
which, through $m_\pi^2\sim m_q$ translates into a dependence on the pion 
mass. The chiral limit of the nucleon mass and of the sigma 
term, respectively, has been evaluated up to NNLO \cite{weise04}, 
where the corrections to the NLO dependence were, however, found to be small. 

In order to be consistent with the $NN$ interaction we account in the 
following for the pion mass dependence of the nucleon mass at 
NLO (expressions given in \cite{weise04}) when the self-energy components 
and the EOS are studied in the chiral limit.

%%%%%%%%%%%%%%%%%%%%%%%%%%%%%%%%%%%%%%%%%%%
\section{Nuclear matter in the chiral limit}
%%%%%%%%%%%%%%%%%%%%%%%%%%%%%%%%%%%%%%%%%%%

The analysis of the EFT potential performed in \cite{plohl06_2} revealed 
that the scalar/vector fields are generated  
by contact terms which occur at NLO order in the chiral 
expansion. These are four-nucleon contact terms with two derivatives 
which generate the short-range spin-orbit interaction. The strength 
of the corresponding low energy constants, in particular those connected 
to the spin-orbit force, is dictated by $P$-wave $NN$ scattering data.  
Pion dynamics as well as LO and N$^3$LO contacts provide 
only corrections to the fields generated by the NLO contact terms. 
Thus one could expect that the quark mass dependence of the fields 
is mainly determined by the quark mass dependence of the contact 
terms which is moderate. 
%%%%%%%%%%%%%%%%%%%%%%%%%%%%%%%%%%%%%%%%%%%%%%%%%%%%%%%%%%%%%%%%%%%%%%%%
\begin{figure}[t]
\includegraphics[width=1\textwidth] {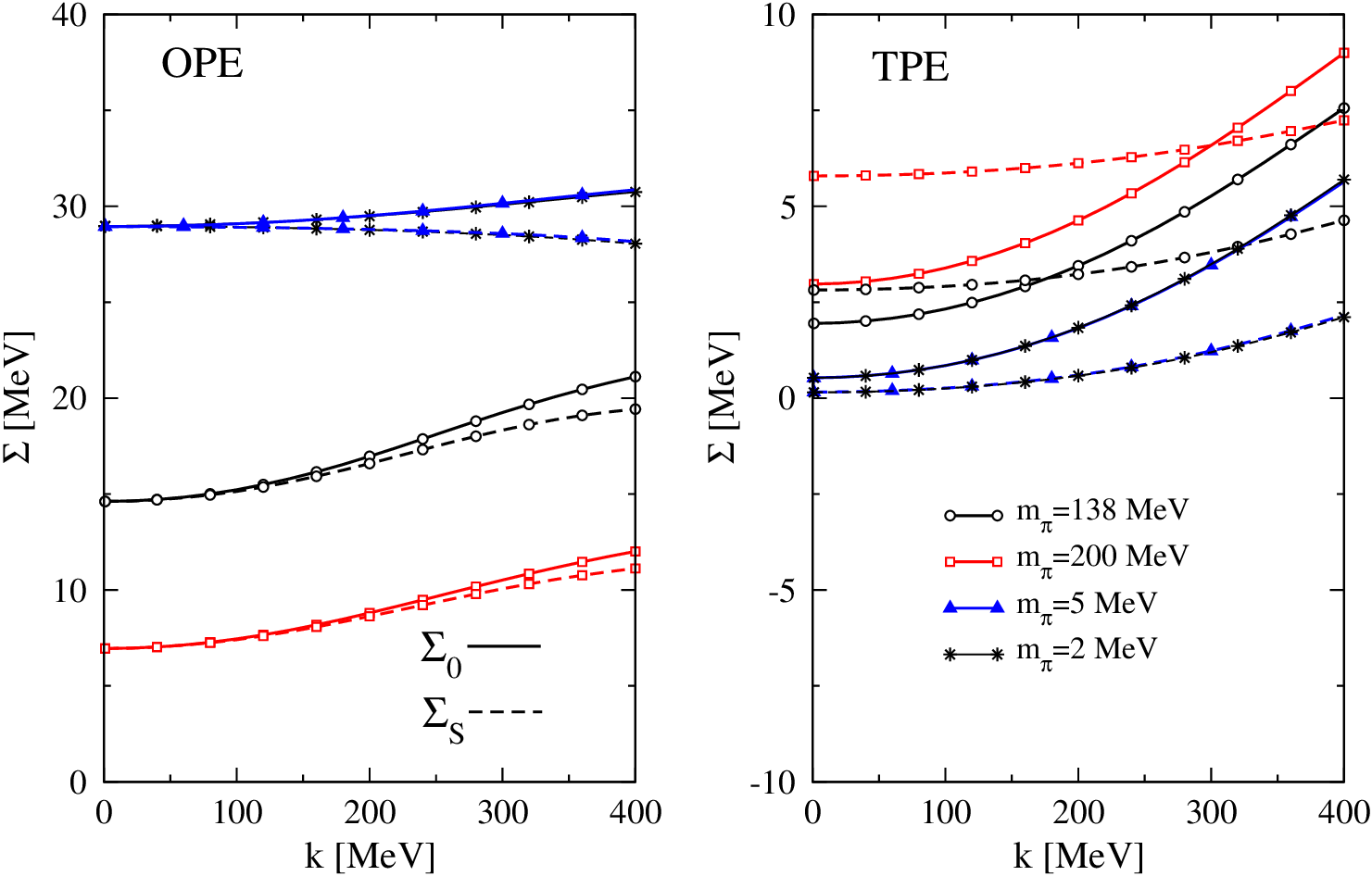}
\caption {(Color online) 
The tree level scalar (dashed lines) and vector (full lines) self-energy 
components in matter at $\kf=1.35~{\rm fm}^{-1}$ are shown for different values 
of the pion mass $m_\pi$. The NLO one-pion exchange (left panel) and 
two-pion exchange (right panel) are shown. The one-pion exchange is obtained
with  $\bar d_{16}=-1.23$ and $\bar d_{18}=-0.97$.
\label{O_T_PE} }
\end{figure}
%%%%%%%%%%%%%%%%%%%%%%%%%%%%%%%%%%%%%%%%%%%%%%%%%%%%%%%%%%%%%%%%%%%%%%%%
However, before coming to the full self-energy, we consider the contributions 
from one-pion (OPE) and two-pion-exchange (TPE) separately. 
Fig.~\ref{O_T_PE} shows the scalar $\Sigs$ and vector $\Sigo$ 
self-energy components from the next-to-leading order OPE and TPE 
contributions at a Fermi momentum of $\kf=1.35~{\rm fm}^{-1}$ which 
corresponds to a nuclear density of $\rho_B=0.166~{\rm fm}^{-3}$. 
As a well known result, at the physical pion mass the scalar and vector 
self-energy 
components from the pseudo-vector OPE are of the same sign and of moderate 
strength. This is also true in the chiral limit. The self-energy components
$\Sigs$, $\Sigo$ approach a constant value of about 30 MeV. 
For the not yet uniquely fixed LECs entering into the expression for the 
renormalized OPE the values $\bar d_{16}=-1.23$ and $\bar d_{18}=-0.97$ have  
been taken~\cite{epelbaum03}.
The uncertainty due to these LECs $\bar d_{16,18}$ does not significantly affect 
the scalar and vector fields. The same is true for the corresponding EOS (see below). 
This is, however,  not the case what concerns the scalar quark 
condensate as discussed in detail in the following chapter.

The scalar $\Sigs$ and vector $\Sigo$ 
self-energy components generated by the TPE are already small for the physical 
case ($m_\pi=138~{\rm MeV}$) and are further reduced by $\approx 2,5$ MeV 
in the chiral limit. At zero momentum both components almost vanish and show 
a slight increase with increasing momentum. 
As for the OPE the fields are repulsive and 
approach a constant value in the chiral limit.
%%%%%%%%%%%%%%%%%%%%%%%%%%%%%%%%%%%%%%%%%%%%%%%%%%%%%%%%%%%%%%%%%%%%%%%%
\begin{figure*}[t]
\includegraphics[width=1\textwidth] {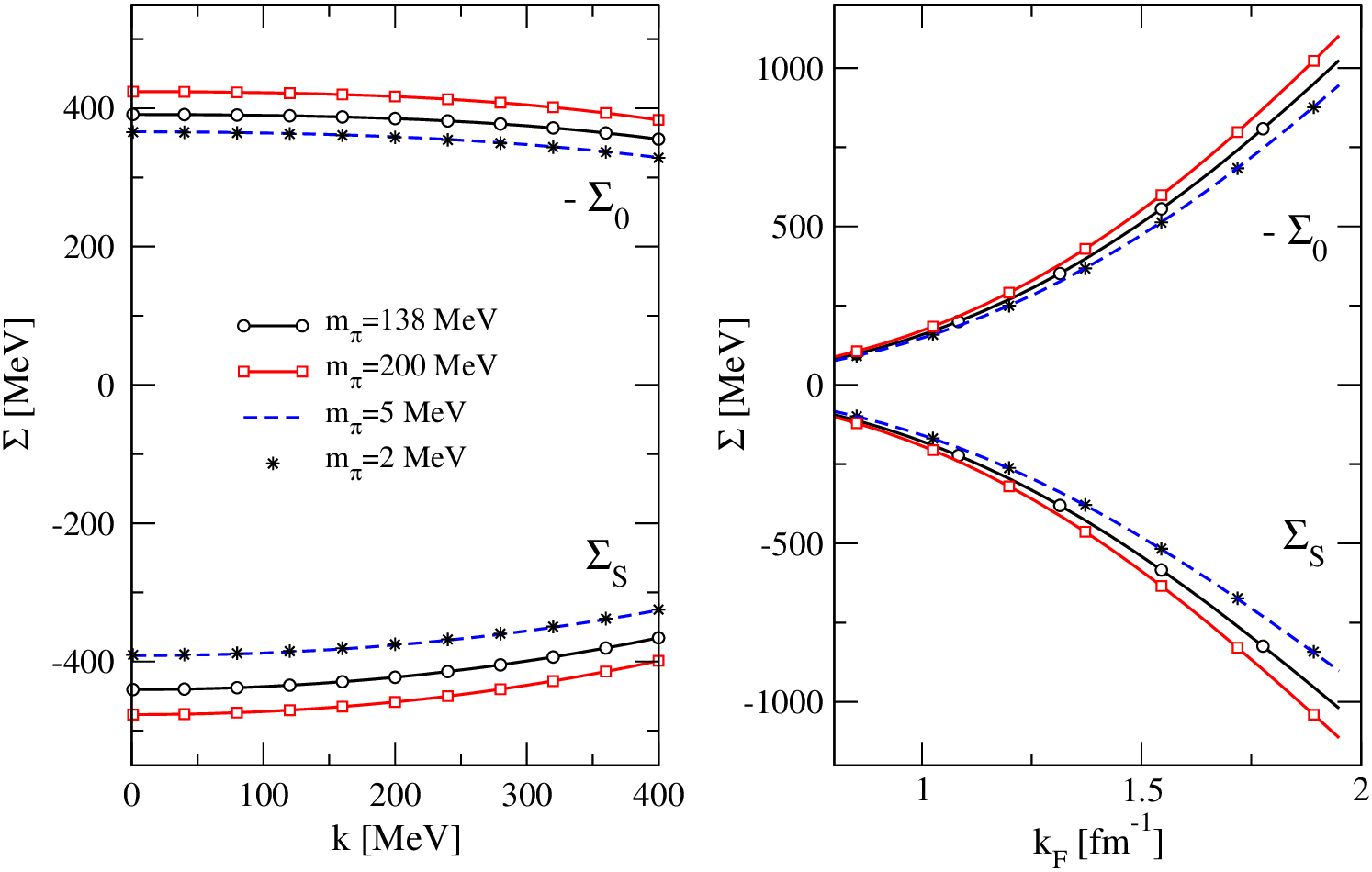}
\caption {(Color online) Left: momentum dependence of the 
tree level scalar and vector self-energy 
components in nuclear matter at $\kf =1.35~{\rm fm}^{-1}$ evaluated for 
different values of the pion mass $m_\pi$.
Right: tree level scalar and vector self-energy components 
in nuclear matter as a function of the Fermi momentum $\kf$ for different 
values of the pion mass $m_\pi$. 
\label{sigma_fig}}
\end{figure*}
%%%%%%%%%%%%%%%%%%%%%%%%%%%%%%%%%%%%%%%%%%%%%%%%%%%%%%%%%%%%%%%%%%%%%%%%

Next we will consider the role of the contact terms. 
The contact terms connected to the LECs $C_{1\ldots 7}$, 
Eq.~(\ref{potfincont}), do not depend on the pion mass at NLO. Since 
the magnitude of the scalar and vector self-energy components is mainly set 
by contact
interactions connected to the spin-orbit force where the strength 
is proportional to the LEC $C_5$ in Eq.~(\ref{potfincont}) the modification 
of the fields in the chiral limit can in total be expected to be moderate.  
The pion mass dependent part of the contact 
interactions,~i.e., the first to lines in Eq.~(\ref{potfincont})
provides only small contributions.

The uncertainties due to the not yet uniquely 
fixed LECs $\bar d_{16,18}$ entering the renormalized OPE (\ref{potfin})
do not strongly affect the self-energy components. However, a second source 
 of uncertainty appears in the the part of the contact interactions 
connected to the not known 
LECs $\bar D_{S,T}$ which depend on the pion mass. In Ref. \cite{epelbaum03} 
this range of uncertainty has been explored through an independent 
variation of the parameters $\alpha_{S,T}$ in Eq.~(\ref{DbarST}) in 
the range of $-3.0 < \alpha_{S,T} < 3.0$.  In Ref. \cite{epelbaum03} 
this rather wide variation of the LECs 
$\bar D_{S,T}$ was motivated by a wide range
of possible parameter sets of NLO LECs fitted with different cut-off 
combinations. However, in the present case - using the Idaho chiral 
potential - we are restricted to one parameter set with a general  cut-off of 
$\lambda=500$ MeV.  A variation of the LECs over a wide range is therefore 
likely to overestimate the uncertainty originating from the LECs $\bar D_{S,T}$. 
Therefore we restrict the present discussion to values 
$\alpha_{S,T} \approx 1$. 
Results turned out to be stable against a variation of 
$\alpha_{S}$ and $\alpha_{T}$ 
in the same direction, i.e. small deviations from combinations 
of $\alpha_{S,T}$ where both parameters are close to each other do 
practically not change the results. 

In Fig.~\ref{sigma_fig} the full tree-level self-energy components 
are now shown as a function of the 
momentum $k$. 
An approximately vanishing pion mass,~i.e. $m_\pi=2$ MeV and $m_\pi=5$ MeV, 
leads to a small reduction of the 
repulsive vector field ($\approx 30$ MeV) 
and of the attractive scalar field ($\approx 50$ MeV), respectively.
The same can be seen on the right hand side in 
Fig.~\ref{sigma_fig} where the fields are shown as a function of the Fermi 
momentum $\kf$. An increase of the pion mass to $m_\pi=200 $ MeV leads to 
an opposite behavior.

In summary, a careful analysis of the chiral EFT $NN$ interaction 
leads to large scalar and vector fields which essentially 
maintain their strength in the chiral limit, however, with the tendency of 
a slight decrease of absolute size.

In this context it may be interesting to compare this behavior with 
the naive assumption of a dropping $\sigma$ meson 
mass within 
the framework of  Quantum-Hadron-Dynamics (QHD) \cite{sw86}. In this case 
the scalar and vector fields are inverse proportional to the masses of 
the $\sigma$ and $\omega$ mesons
\beqa
\Sigs = -\frac{g_\sigma^2}{m_\sigma^2}\rho_S \quad ,\quad 
\Sigo = +\frac{g_\omega^2}{m_\omega^2}\rho_B
\label{qhd1}
\eeqa
where $g_\sigma$ and  $g_\omega$ are the corresponding meson-nucleon 
coupling constants and $\rho_S \sim \rho_B$ is the scalar nucleon density. 
The assumption of dropping $\sigma$ and $\omega$ meson masses according to a naive 
Brown-Rho scaling~\cite{brown96} together with fairly constant couplings would 
lead to a strong increase of scalar and vector fields in size. Chiral EFT 
predicts the opposite behavior, namely slightly decreasing fields. Interpreting 
this result in terms of the simple QHD picture means that the ratio of 
coupling functions and meson masses in Eq. (\ref{qhd1}) has to stay fairly 
constant. Assuming dropping meson masses the coupling functions should show 
the same density dependence. Such a scenario is not completely unrealistic since 
in the framework of density dependent relativistic mean field theory \cite{Fuc95} 
where density dependent meson coupling functions $g^2_{\sigma,\omega}(\rho_B)$ are 
derived from the Brueckner G-matrix~\cite{Hof01,dalen06} or 
fitted to finite nuclei~\cite{Typ99,Nik02}, such a behavior is usually 
obtained.

%%%%%%%%%%%%%%%%%%%%%%%%%%%%%%%%%%%%%%%%%%%%%%%%%%%%%%%%%%%%%%%%%%%%%%%%
\begin{figure*}[t]
\includegraphics[width=1\textwidth] {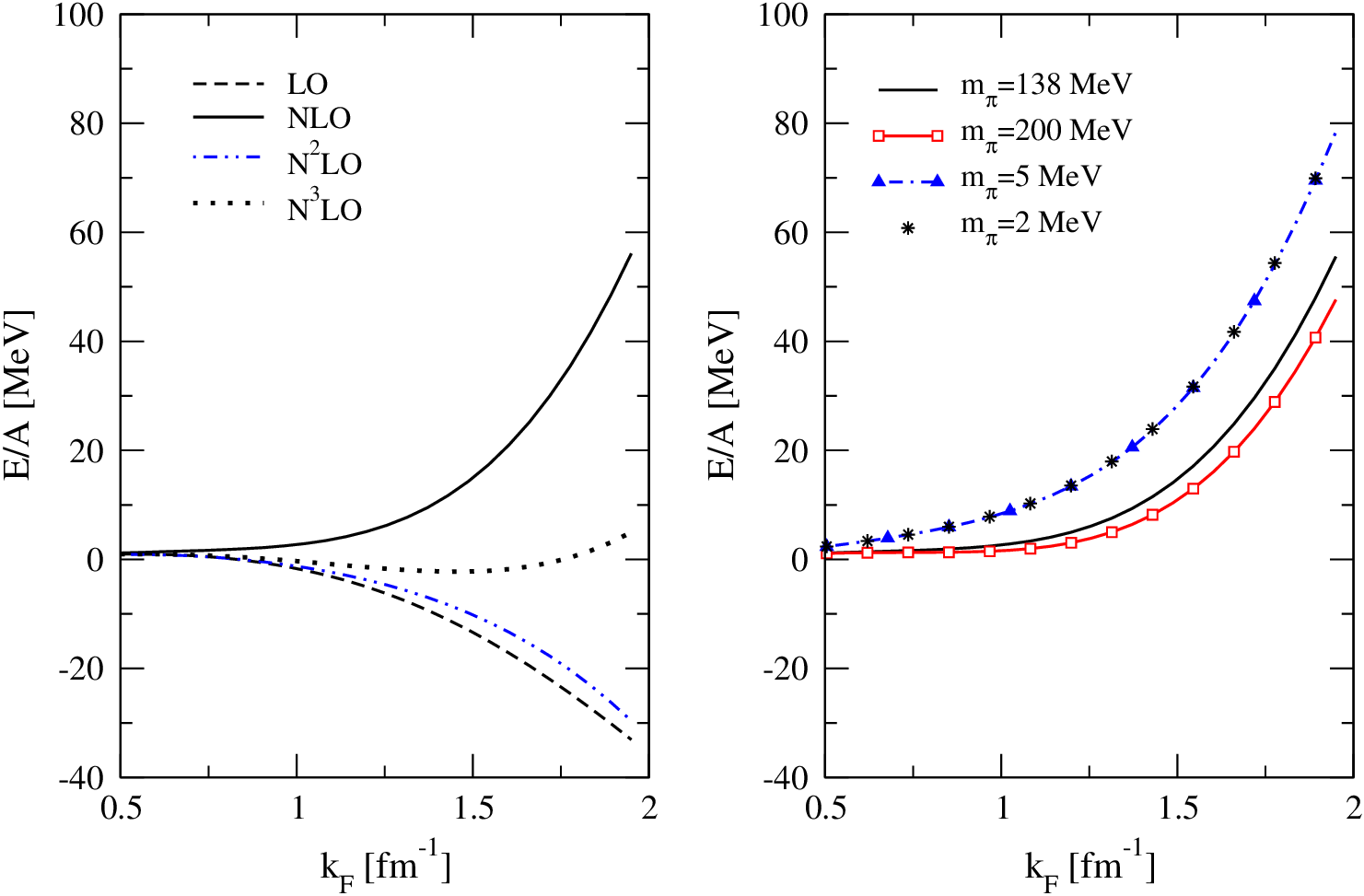}
\caption {(Color online) Hartree-Fock calculation of the nuclear equation of 
state, i.e. the energy per particle $E/A$ as a function of the 
Fermi momentum $k_F$.
On the left hand side the tree level results are shown order by order up to N$^3$LO.
On the right hand side the pion mass dependence of the EOS at NLO is 
shown again, i.e. the result for the physical
case of $m_\pi=138$ MeV is compared to the case of $m_\pi=2$ MeV, 
$m_\pi=5$ MeV and 
$m_\pi=200$ MeV.
\label{eos_orders} }
\end{figure*}
%%%%%%%%%%%%%%%%%%%%%%%%%%%%%%%%%%%%%%%%%%%%%%%%%%%%%%%%%%%%%%%%%%%%%%%%

It is clear that a rather small reduction of the scalar and vector fields 
has only moderate consequences for the EOS in the chiral limit.  
How the change in the fields affects the nuclear EOS is 
depicted in Fig. \ref{eos_orders}. However, before coming to the pion 
mass dependence we shortly discuss the tree level EOS 
derived from the chiral EFT potential. The EOSs for isospin symmetric nuclear
matter at the various orders of the potential are shown on the left hand side
of Fig. \ref{eos_orders}. There appear large jumps in the 
EOS when going from LO up to N$^3$LO. As discussed in \cite{plohl06_2} the 
contact terms which generate the large attractive/repulsive scalar/vector 
potentials arise at NLO. Contributions from higher order provide corrections 
to these potentials. These are moderate on the scale of the fields of 
several hundred MeV magnitude. However, due to the subtle 
cancellation between scalar attraction and vector repulsion such corrections 
may be large on the scale of the binding energy, i.e. several tenth of MeV. 
This behavior is exactly reflected in the EOSs shown in Fig. \ref{eos_orders} 
which jump from unbound at LO to over-bound at NLO and N$^2$LO to loosely 
bound at   N$^3$LO. 

The present tree level calculation is of course not a realistic 
microscopic nuclear matter calculation which would on the one hand 
require to apply the  N$^3$LO force, and secondly, to perform a self-consistent 
summation of the Brueckner ladder diagrams \cite{gross99,dalen04}. 
The difference 
between a tree-level and a full relativistic Brueckner calculation 
has been discussed in~\cite{plohl06_2}. 
However, the NLO tree level calculation allows a 
consistent investigation of the chiral limit at the order at which 
the pion mass dependence of the chiral $NN$ potential has been derived.

This is done in the left panel of Fig.~\ref{eos_orders} which 
compares the  NLO EOS at the physical pion mass $m_\pi=138$ MeV to 
that at  $m_\pi=200$ MeV, $m_\pi=2$ MeV and $m_\pi=5$ MeV, respectively. 
Doing so, one observes a slight softening of the 
repulsive NLO EOS when the pion mass is increased from its physical value 
to $m_\pi=200$ MeV. In the chiral limit the scalar attraction is slightly 
stronger reduced than the vector repulsion 
(see Fig.~\ref{sigma_fig}). The EOS becomes therefore more repulsive. 
One can expect that the observed effect survives also when a full Brueckner ladder 
is summed. Since the pion mass dependence of the contact terms is rather 
weak, Brueckner short range correlations can not be expected to 
change the results dramatically. As already pointed out, 
the main source of uncertainty arises from the unknown LECs $\bar D_{S,T}$ 
entering the renormalized contact forces. Concerning the tree level 
EOS, we did, however, 
not find any tendency of a qualitative change even for large, 
probably unrealistic variations of the dimensionless
coefficients $\alpha_{S,T}$. In a full Brueckner calculation 
the attraction may be increased  since iterated 
OPE and TPE is known to be quenched by Pauli blocking of the intermediate 
states in the Bethe-Goldstone, respectively the Bethe-Salpeter equation.

Therefore the conclusion is, that 
the magnitude of the large scalar and vector fields in matter persists 
in the chiral limit and that the physics of infinite nuclear matter is 
similar to
that of a vanishing pion mass. This finding is in agreement with the 
investigations made in~\cite{bulgac56}.
In this work nuclear matter was analyzed taking basically
the chiral limit of the OPE and assuming that the short range and 
the intermediate range part of any $NN$ potential is not affected.

%%%%%%%%%%%%%%%%%%%%%%%%%%%%%%%%%%%%%%%%%%%
\section{In-medium scalar condensate}
%%%%%%%%%%%%%%%%%%%%%%%%%%%%%%%%%%%%%%%%%%%

The spontaneous breakdown of chiral symmetry involves a qualitative 
rearrangement of the QCD ground state, due to the appearance of scalar 
quark-antiquark pairs. The corresponding non-vanishing ground-state 
expectation value $\langle  {\bar q}q\rangle$, denoted as
the scalar quark condensate, is an order parameter  of spontaneous chiral 
symmetry breaking. Any reduction of the scalar density of quarks
in matter can therefore be interpreted as a sign of partial restoration of 
chiral symmetry. 

%%%%%%%%%%%%%%%%%%%%%%%%%%%%%%%%%%%%%%%%%%%
\subsection{Determination from Hellmann-Feynman theorem}
%%%%%%%%%%%%%%%%%%%%%%%%%%%%%%%%%%%%%%%%%%%

The vacuum value of the lowest-dimensional quark  condensate is about~\cite{gasser82}  
\begin{equation}
\langle {\bar q} q\rangle_0\simeq -(225\pm25\,\mathrm{MeV} )^3~~.
\end{equation}
The density dependence of the chiral condensate 
$\langle {\bar q}q \rangle$ can be extracted exploiting 
the Hellmann-Feynman theorem with respect
 to the symmetry breaking current quark mass term of the 
QCD Hamiltonian. We consider isospin symmetric matter making thereby 
use of the isospin symmetry of the condensates 
($\langle {\bar q}q \rangle \equiv \langle {\bar u}u 
\rangle\simeq \langle {\bar d}d \rangle $). 
Defining $\bar q q\equiv\frac{1}{2}(\bar u u+ \bar d d)$ and 
$m_q \equiv \frac{1}{2}(m_u+m_d)$ the quark mass term is given by 
$2m_q{\bar q} q$. Isospin-breaking terms are neglected. 
With the help of the Hellmann-Feynman theorem one obtains the in-medium quark 
condensate by determining the energy density $\mathcal{E}$ of nuclear matter
\begin{equation}
  2m_q(\langle \bar q q\rangle_{\rho_B}-\langle \bar q q\rangle_0)=m_q\frac{d\mathcal{E}}{dm_q}.
\label{HF1}
\end{equation}
The derivative is taken at fixed density.
The energy density of nuclear matter is given by 
\begin{equation}
  \mathcal{E}=M\rho_B+\frac{E(\rho_B)}{A}\rho_B
\label{energydens}
\end{equation}
where the second term of $\mathcal{E}$ is the energy per particle $E/A$ (times 
the baryon density), i.e. 
the contributions from the nucleon kinetic energy and nucleon-nucleon 
interactions.
Inserting the energy density $\mathcal{E}$ into Eq.~(\ref{HF1}) and using the 
Gell-Mann, Oakes, Renner (GOR) relation 
\begin{equation}
2m_q\langle\bar q q\rangle_0=-m_\pi^2f_\pi^2
\label{GOR}
\end{equation}
with the definition of the pion-nucleon $\sigma_N$ term from Eq.~(\ref{sigma1}) 
\begin{equation}
\frac{dM}{dm_q}=\frac{\sigma_N}{m_q}
\end{equation}
one obtains
\begin{equation}
\frac{\langle \bar q q\rangle_{\rho_B}}{\langle \bar q q\rangle_0}=
1-\frac{\rho_B}{m_\pi^2f_\pi^2}\left[\sigma_N+m_q\frac{d}{dm_q} 
\frac{E}{A}\right].
\label{HF2}
\end{equation}
The derivative of the energy per particle with respect to the quark mass can 
be re-expressed using the chain rule
\begin{equation}
\frac{\langle \bar q q\rangle_{\rho_B}}{\langle \bar q q\rangle_0}=
1-\frac{\rho_B}{m_\pi^2f_\pi^2}\left[\sigma_N+m_q\frac{\partial (E/A)}
{\partial M}\frac{dM}{dm_q}+
m_q\frac{\partial (E/A)}{\partial m_\pi}\frac{dm_\pi}{dm_q} \right].
\end{equation}
The derivative of the pion mass using Eq.~(\ref{GOR}) is given by
\begin{equation}
\frac{dm_\pi}{dm_q}=\frac{m_\pi}{2m_q}
\end{equation}
valid to leading order in chiral perturbation theory. Introducing 
\begin{equation}
\rho^\chi=\frac{m^2_\pi f^2_\pi}{\sigma_N}
\end{equation}
one finally obtains
\begin{equation}
\frac{\langle \bar q q\rangle_{\rho_B}}{\langle \bar q q\rangle_0}=
1-\frac{\rho_B}{\rho^{\chi}}\left[1+\frac{\partial (E/A)}{\partial M}+
\frac{\partial (E/A)}{\partial m_\pi}\frac{m_\pi}{2\sigma_N}\right].
\label{HF3}
\end{equation}
The first term on the r.h.s. of Eq.~(\ref{HF3}) which reduces the 
condensate in matter is model independent and of first order in the nuclear 
density ~\cite{cohen91,drukarev91}. 
Inserting  the empirical value   of  $\sigma_N =( 45\pm 7)$ MeV 
for the sigma term ~\cite{gasser91} and taking $m_\pi=138$ MeV and $f_\pi=92.4$ MeV 
one finds in Table~\ref{tab1}, 
that the in-medium scalar condensate is to leading order in density
approximately $\frac{1}{3}$ smaller than its vacuum value at nuclear saturation density. 
In the following a value of  $\rho_0=0.173$ fm$^{-3}$, 
corresponding to a Fermi momentum of 
$k_F=1.37$ fm$^{-1}$, is chosen as the standard value for the nuclear
saturation density.

\begin{figure*}[t]
\includegraphics[width=1\textwidth] {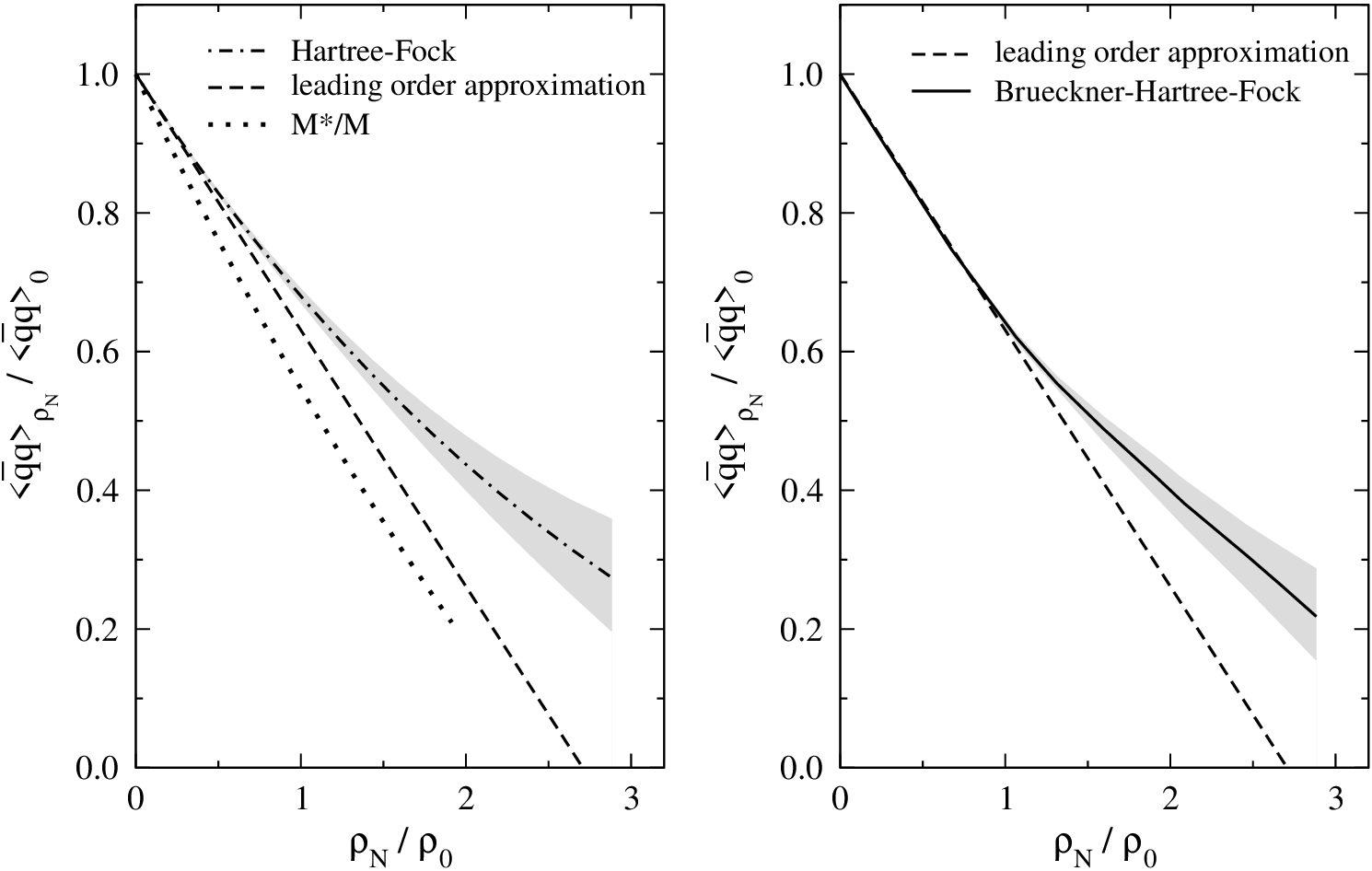}
\caption {In-medium scalar quark condensate as a function of 
 density with $\sigma_N=45$~MeV obtained in various approximations:  
A  Hartree-Fock calculation of 
$\langle \bar q q\rangle_{\rho_B}/\langle \bar q q\rangle_0$
is shown compared to $M^*/M$ where $M^*=M+\Sigs$ is evaluated at tree 
level in Hartree-Fock approximation (left).
A full calculation based on the Brueckner-Hartree-Fock approach  is shown 
on the right.
The light shaded bands indicate  the uncertainty 
of the corresponding result  varying the LECs
$\bar d_{16}$ from $\bar d_{16}=-0.91$ GeV$^{-2}$ to $\bar d_{16}=-1.76$ 
GeV$^{-2}$ and $\bar d_{18}$ from $\bar d_{18}=-0.84$ GeV$^{-2}$ to 
$\bar d_{18}=-1.54$ GeV$^{-2}$.
Dashed line: model-independent leading order result. 
\label{scalarcond} }
\end{figure*}

From Fig.~\ref{scalarcond} one sees that to leading order a 
complete restoration of chiral symmetry would already 
occur at $\rho_B\approx 2.7\rho_0$. Such a scenario is unrealistic and 
contradictory to the knowledge from heavy ion reactions \cite{fuchs06} 
and astrophysics, e.g. neutron stars \cite{klaehn06}. 
Hence, one has to account for higher order corrections 
in density coming from the $d(E/A)/dm_q$ term in Eq.~(\ref{HF2}). 
One might estimate this correction to be small due to the binding energy of  
$E/A\approx -16$ MeV, which is 
almost two orders of magnitude smaller than the nucleon mass 
contributing dominantly to the energy density in Eq.~(\ref{energydens}). 
Nevertheless, since the  quark mass derivative of the interaction energy is 
the relevant quantity, it is by far not obvious that higher order corrections 
are negligible.

Thus, a reliable extraction of the density dependent scalar condensate 
$\langle \bar q q\rangle_{\rho_B}$ requires both, a 
sophisticated nuclear matter calculation {\it and} 
the exact knowledge of the current quark mass dependence of 
{\it all} model parameters entering into the energy density. 

Previous estimates of the scalar 
condensate based on sophisticated ab-initio many-body 
approaches~ \cite{brockmann96,Li94} 
suffered from this problem. The relativistic Brueckner approach 
chosen in \cite{brockmann96,Li94} provides a reliable description of 
nuclear matter bulk properties. Such calculations are based  
on realistic $NN$ potentials, e.g. one-boson exchange 
potentials~\cite{bonn}. However, the current quark mass dependences of 
the model parameters, i.e.  meson masses and coupling constants, are unknown 
to large extent and have therefore either been roughly estimated or 
even been neglected~\cite{brockmann96,Li94}. 

In ~\cite{brockmann96} it was found  that the largest 
uncertainty in the calculation of the in-medium quark condensate 
$\langle \bar q q\rangle_{\rho_B}$ arises due to the unknown quark mass 
dependence of the scalar isoscalar $\sigma$ meson exchange which 
parameterizes effectively correlated two-pion exchange.
 
In the present work the determination of the energy per particle 
$E/A$ will be determined within Hartree-Fock (HF), Eq.~(\ref{eos1}),  
and in a second step within the Brueckner-Hartree-Fock (BHF) approximation. 
Since the $NN$ interaction is thereby based on chiral EFT, 
Eqs.~(\ref{potfin})-(\ref{potfincont}), where the complete 
pion mass dependence is known up to NLO ~\cite{epelbaum03} we are free 
of uncertainties concerning unknown quark mass derivatives. 
Remaining ambiguities when applying the 
Hellman-Feynman theorem are only due to the not yet uniquely 
fixed LECs $\bar D_{S,T}$ in the NLO contact terms, see 
Eq.~(\ref{potfincont}), and the LECs $\bar d_{16,18}$ showing up in the OPE 
exchange. The uncertainties coming from these LECs will be discussed. 

In Hartree-Fock approximation nuclear matter is normally unbound, 
in particular when high precision OBE type potentials are applied. 
The situation turns out to be qualitatively different 
for low momentum interactions like $V_{\rm low~k}$ and Idaho N$^3$LO 
(see Fig. \ref{eos_orders})  where the hard 
core is strongly suppressed by high momentum cut-offs. In this case isospin 
saturated nuclear matter collapses and Brueckner ladder correlations do 
not improve on~\cite{kuckei02,bogner05}. 
In this case the repulsion 
generated by three-body forces which appear at N$^2$LO turned out to 
be essential to stabilize nuclear matter and to obtain reasonable 
saturation properties \cite{bogner05}. The inclusion of three-body forces 
is beyond the scope of the present work, in particular what concerns the 
pion mass dependence, but for a quantitative determination of the EFT EOS 
they should be included.

However, for a reliable estimate of 
the in-medium condensate the role of $NN$ correlations,~in particular 
short-range and tensor correlations, has to be considerd. One 
might assume that $NN$ correlations influence the 
result for the condensate, in particular at higher densities.
In order to estimate their importance the self-consistent 
iteration scheme of BHF theory is applied.

The central equation of the BHF approximation is the
Bethe-Goldstone equation
\beq
\mathcal{G} (\omega)= V+ V \frac{Q}{\omega-H_0}\mathcal{G}(\omega).
\eeq
$V$ is the bare interaction and $Q$ the Pauli operator which prevents from   
scattering into  occupied intermediate 
states below the Fermi momentum $k_F$. The starting energy is denoted by 
$\omega$. The operator $H_0$ defines the energy spectrum of the intermediate
two-particle state where we use the so called continuous choice.
Therefore the single-particle energies for particles as well as for  hole 
states above the Fermi
surface are calculated from the kinetic energy and a mean field part which has 
to be determined from the $\mathcal G$ matrix self-consistently.
The single-particle energies are then given by
\beq
\epsilon_{\alpha}^{BHF}=\epsilon_{\alpha}+\sum_{\mu \le F} \langle \alpha \mu 
|\mathcal{G} (\omega=\epsilon_{\alpha}^{BHF}+\epsilon_{\mu}^{BHF})|\alpha \mu \rangle.
\eeq
These single-particle energies are then parameterized in terms of an effective 
mass and a constant potential,~i.e. $\epsilon_{k}\approx k^2/2M^*+U$.

In the present work we apply the non-relativistic 
approach since it is not possible to use the chiral $NN$ potential in a 
relativistic BHF calculation (DHBF) \cite{gross99,dalen04} where 
one accounts in addition for the dressing of the potential matrix 
elements $V\mapsto V^*$. The latter requires, however, a 
definite relativistic structure of the interaction, like for 
covariant OBE-type potentials. Nevertheless, differences between 
a relativistic and a non-relativistic
treatment should  be moderate concerning the derivative of the 
EOS with respect to the current quark mass.
%%%%%%%%%%%%%%%%%%%%%%%%%%%%%%%%%%%%%%%%%%%%%%%%%
%%%%%%%%%%%%%%%%%%%%%%%%%%%%%%%%%%%%%%%%%%%%%%%%%

The prediction of the in-medium scalar condensate in both approaches,~i.e., 
HF and BHF are shown in Fig.~\ref{scalarcond}. 
As expected, deviations from the leading order result 
due to $NN$ interactions and nucleon kinetic
energy, Eq.~(\ref{HF3}), increase with density. 
For both approaches, HF and BHF, the additional contributions 
lead to a weaker reduction of the in-medium quark 
condensate. Especially in the case of the BHF calculation the leading order
prediction provides a very good description of the quark condensate 
up to a density of $0.8\rho_B$. 

At nuclear saturation density $\rho_0$ the reduction of the in-medium 
quark condensate is about $3\%$ (BHF) and  $12\%$ (HF) 
smaller compared to leading order. Deviations are, however, 
growing with density, where at $\rho_B \approx 2\div 3 \rho_0$ the quark condensate 
is reduced to $\approx 35 \%$ (HF) and $\approx 30 \%$ (BHF).
Naturally the BHF approach is more reliable in this density region. 
However, in summary effects from 
short-range $NN$ correlations and the quenching of OPE and TPE due to 
Pauli blocking, both present in BHF, have only minor implications 
for the condensate as can be seen from the comparison to the HF 
result.

The uncertainty due to the not yet uniquely fixed LECs $\bar d_{16,18}$ 
in the renormalized OPE,~Eq.~(\ref{potfin}) which was already 
mentioned in the context of the EOS in the chiral limit, enters also 
into the determination of the scalar condensate. However, now this 
uncertainty is much more severe. 
The light shaded bands in Fig.~\ref{scalarcond} indicate the range of possible
variations: The LEC $\bar d_{18}$ is extracted from the Goldberger-Treiman discrepancy.
We take the three empirically found values given in~\cite{epelbaum03}
extracted from three different phase shift, 
$\bar d_{18}=-0.84$ GeV$^{-2}$~\cite{said}, 
$\bar d_{18}=-0.97$ GeV$^{-2}$~\cite{matsinos} and 
$\bar d_{18}=-1.54$ GeV$^{-2}$~\cite{koch86}, respectively.
Furthermore we vary the LEC $\bar d_{16}$ in the range from 
$\bar d_{16}=-0.91$ to $\bar d_{16}=-1.76$ as done in~\cite{epelbaum03}.
The upper bound of the shaded band  corresponds to $\bar d_{16}=-1.76$ and
$\bar d_{18}=-0.84$ whereas the lower bound corresponds to $\bar d_{16}=-0.91$
and $\bar d_{18}=-1.54$. 
These  uncertainties are also given in Table~\ref{tab1}. 
The HF (dash-dotted line) and BHF (solid line) mean values are 
obtained with  $\bar d_{16}=-1.23$ and $\bar d_{18}=-0.97$.

%%%%%%%%%%%%%%%%%%
\newcolumntype{Z}{>{\centering\arraybackslash}X}
\begin{table}
\caption{\label{tab1} Predictions of $\langle \bar q q\rangle_{\rho_B}/\langle \bar q q\rangle_0$ obtained with the Hellmann-Feynman theorem in diverse 
approximations compared with $M^*/M$ for three different 
values of the nucleon density $\rho_B$.}
\begin{center}
\begin{tabularx}{\textwidth}{ZZZZZ}
\hline\hline
&  \multicolumn{3}{c}{\bf Hellmann-Feynman theorem}& { $\bf M^*/M$}\\ 
$\rho_B/\rho_0 $  & Leading order  & HF  & BHF& in HF \\ 
\hline
0.5  & 0.815 & $0.828\pm 0.002$  & $0.812\pm 0.001$  & 0.759\\
1.0  & 0.630 & $0.677\pm 0.010$  & $0.641\pm 0.004$  & 0.546   \\
1.5  & 0.445 & $0.550\pm 0.020$  & $0.510\pm 0.014$  & 0.354 \\
\hline\hline
\end{tabularx}
\end{center}
\end{table}
%%%%%%%%%%%%%%%%%%
Comparing with previous approaches performed in a similar spirit 
\cite{Li94,brockmann96} we find generally a 
stronger reduction of the scalar condensate. 
In~\cite{Li94} calculations were based on the relativistic DBHF approach and 
an one-boson-exchange potential (Bonn A) has been used. In this approach 
an unexpected increase of the in-medium scalar condensate
at densities above $\rho < 2.5\rho_B$ has been found. The same 
tendency, i.e. an increasing quark condensate at high density 
has been observed in~\cite{brockmann96}.
In~\cite{Li94} it was assumed that this increase is caused by  a 
breakdown of the underlying assumptions related to the current quark 
mass dependences of the model parameters, i.e. meson masses and coupling 
constants. 
The authors concluded that the use of not chirally invariant 
$NN$ potentials may lead to wrong predictions in a density region 
where chiral restoration is expected to occur.

As already mentioned we do not face such problems since 
the chirally invariant EFT interaction used in the present work has 
a well defined quark mass
dependence. The only source of uncertainty arise due to the LECs 
$\bar d_{16,18}$ which are not yet uniquely fixed and the unknown 
LECs  $\bar D_{S,T}$ entering the short-range part,~i.e., the contact force
which could provide substantial corrections to the scalar quark condensate. 
Nevertheless, the same argument given in the previous section kept us from 
showing a wide undefined variation. 
Therefore, we restrict our calculation again to the case of 
$\alpha_{S,T}\approx 1$. However, as for the EOS, 
the prediction of the quark condensate is not considerably altered varying 
$\alpha_{S,T}$ for combinations of $\alpha_{S,T}$ where both parameters 
are close to each other. The contributions which change the condensate 
originate  then mainly from  TPE and renormalized contact forces.   
The LECs $C_{1\ldots 7}$ in Eq.~(\ref{potfincont}) do 
not depend on the pion mass after renormalization and 
the related contact terms do therefore not 
contribute to the change of the in-medium quark condensate.

Nevertheless, considering the possible band of variation due to the 
LECs $\bar d_{16,18}$, both, the HF and BHF calculations 
shown in Fig.~\ref{scalarcond} do not indicate 
saturation or even an increase of the condensate in the considered 
density range up to $3\rho_B$. Extrapolating the BHF prediction 
to high densities a complete restoration of 
chiral symmetry,~i.e., a vanishing scalar quark condensate is not likely to
happen below $4\rho_0$, even if one takes the range of uncertainty from 
not yet exactly known LECs into account. 

\begin{figure*}[t]
\includegraphics[width=1\textwidth] {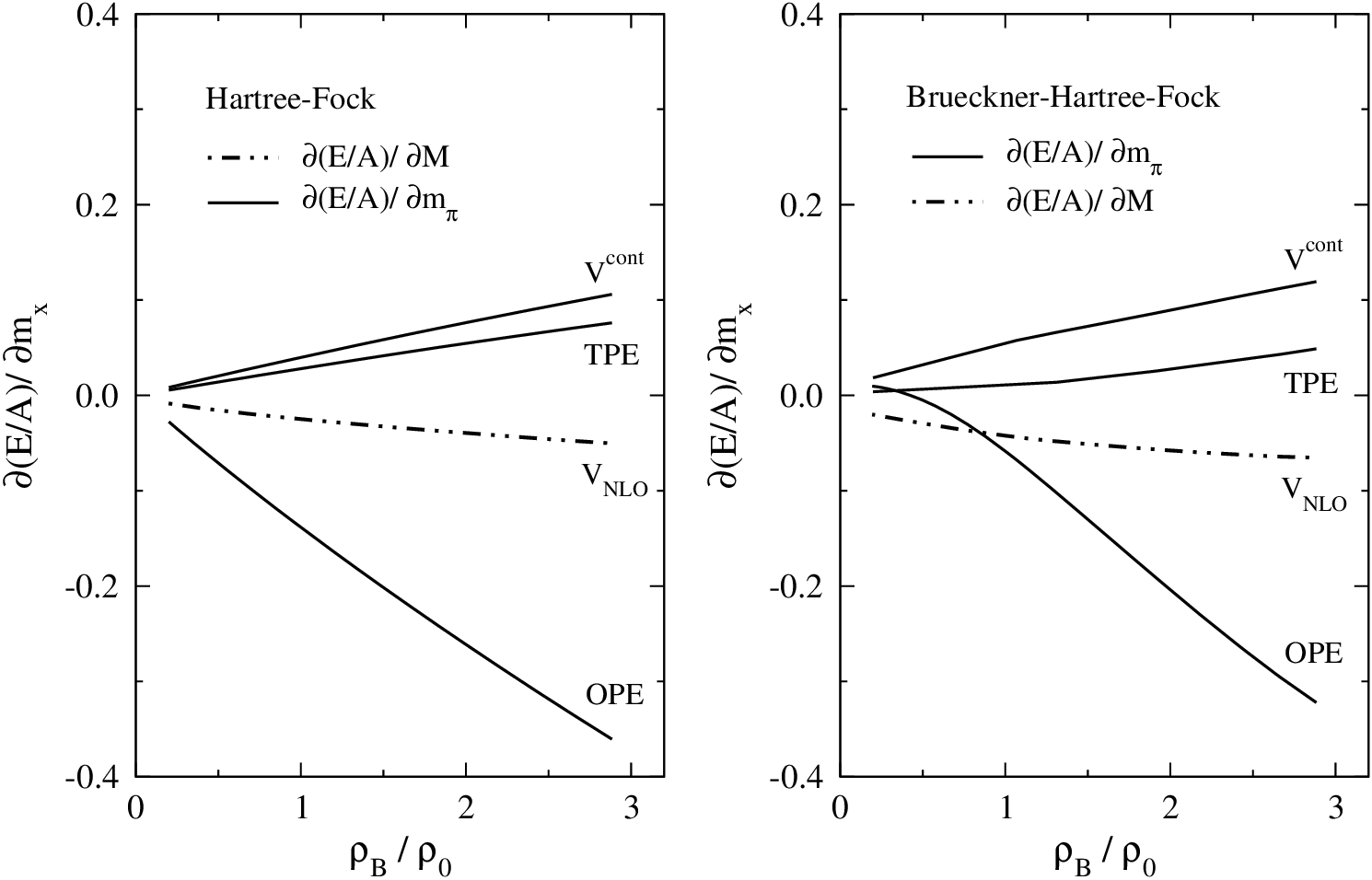}
\caption {The derivatives $\partial (E/A)/\partial m_\pi$ 
as a function of density with respect to the full NLO
calculation as well as the separate contributions, i.e. OPE, TPE and contact 
interactions are shown. Additionally the dash-dotted line denotes
$\partial (E/A)/\partial M$ for the full NLO calculation.
\label{dEdmx} }
\end{figure*}

The first determination of the in-medium quark condensate 
adopting the Hellmann-Feynman theorem 
has been carried out by Cohen et al.~\cite{cohen91}, based 
on the $\pi$-$N$ Fock term. There the 
condensate was found to be reduced to a value of 0.694 at nuclear saturation 
density $\rho_0$ and 0.58 at $1.5\rho_0$ which is in fair agreement with 
$0.677\pm 0.01$ and $0.550\pm 0.020$ obtained in the present HF  
calculation (third column of Table~\ref{tab1}).
This agreement, is, however, somewhat accidental since we find that 
TPE and contact interactions
(which both have not been included in \cite{cohen91}) reduce the 
contribution from OPE by $\approx 50 \%$. 
Moreover, in~\cite{cohen91} the $dg_{\pi N}/dm_q$ dependence has been neglected and 
a different value for $g_{\pi N}$ has been used. Both 
calculations are, however, comparable in the sense that $NN$ correlations are 
neglected and they are of the same order in the density. 
Moreover, short-range physics due to contact terms, which have been 
neglected in~\cite{cohen91} are found to provide only 
moderate corrections as can be seen from Fig.~\ref{dEdmx}. 

%%%%%%%%%%%%%%%%%%%%%%%%%%%%%%%%%%%%%%%%%
%%%%%%%%%%%%%%%%%%%%%%%%%%%%%%%%%%%%%%%%%

In Fig.~\ref{dEdmx} the derivatives $\partial (E/A)/\partial m_\pi$ from 
Eq.~(\ref{HF3}) as a function of the density
are shown with respect to the different contributions coming from pion 
dynamics,~Eqs.~(\ref{potfin}) and~(\ref{potfinTPE}), and 
from the contact terms ($V^con$, Eq.~(\ref{potfincont}), respectively.
The dashed-dotted line indicates the derivative $\partial (E/A)/\partial M$. 
The contribution coming from OPE 
is negative and considerably larger compared to those from 
TPE and contact interactions. 
In general the contributions from pion dynamics,~i.e., OPE and TPE are 
smaller in BHF due to quenching effects.  
In the case of the contact interactions (\ref{potfincont}),
one has to keep in mind that the short-range terms 
$\tilde m^2_\pi\ln\tilde m_\pi$
show up due to TPE and the renormalization of the leading order contact 
terms by pion loops.

The contribution from nucleon interactions is getting substantially more 
important with increasing density compared to the contribution from the 
nucleon 
kinetic energy (dot-dashed line) which is of order $O(\rho^{5/3})$. 
Thus the nuclear interaction provides important corrections to the 
Fermi gas approximation usually made in QCD sum rule approaches.  

It turns out that the calculations are highly sensitive to 
the pion mass dependence of the pion nucleon coupling constant
$g_{\pi N}$   which has been often  neglected in earlier works. 
This fact can also be seen from the relatively large bands of uncertainty 
in Fig.~\ref{scalarcond} since 
the corresponding LECs enter into the relative shift of 
$g_A/F_\pi$, Eq.~(\ref{deltaCL}), which is connected to $g_{\pi N}$ via the
Goldberger-Treiman relation $g_{\pi N}/M=g_A/F_\pi$.

We conclude that 
the contributions from nucleon interactions to the change of the scalar 
condensate in matter are mainly due to low-momentum virtual pions. In 
contrast to the scalar/vector fields which are generated by NLO contact 
interactions \cite{plohl06} 
the contact terms and short-range
correlations,~i.e., the short-distance physics, seem to play a minor role 
for the change of the in-medium quark condensate.
Nevertheless, for a fully reliable prediction of the in-medium 
quark condensate the little known  LECs $\bar D_{S,T}$ entering the 
NLO contact interactions,~Eq.~(\ref{potfincont}), have to be fixed with 
better precision.

%%%%%%%%%%%%%%%%%%%%%%%%%%%%%%%%%%%%%%%%%%%
\subsection{Effective nucleon mass}
%%%%%%%%%%%%%%%%%%%%%%%%%%%%%%%%%%%%%%%%%%%

QCD in-medium sum rules relate the scalar and vector in-medium  condensates 
$\langle {\bar q}q \rangle_{\rho_B}$  
and $\langle q^\dagger q  \rangle_{\rho_B} $ to the isoscalar scalar and
vector self-energies of a nucleon in matter. Thus the model 
independent leading order result which should be valid at low 
density determines the density dependence of the effective nucleon 
mass $M^*=M+\Sigs$ within the in-medium QCD sum rule approach. 
The scalar and vector fields $\Sigs$ and $\Sigo$ arise naturally from 
the structure of the quark propagator which is proportional to the 
corresponding in-medium quark condensate. 
The quark correlation function which follows from the operator product 
expansion can be written to leading order in terms 
of  scalar $\langle {\bar q}q \rangle_{\rho_B} $ 
and vector condensates $\langle  q^\dagger q \rangle_{\rho_B} $. 
In contrast to the scalar condensate 
to leading order the vector condensate is exactly known. 
It is given by the quark density in the 
nuclear matter rest-frame since the baryon current is conserved, i.e. 
$\langle q^\dagger q \rangle_{\rho_N} = 3/2 \rho_B$.
In \cite{plohl06_2} we compared the  NLO EFT vector self-energy
to the leading order sum rule result.
As in the case of the scalar self-energy  deviations were found to be 
small at moderate densities. 
The next order in the operator product expansion involves four-quark 
operators and combinations  of quark and gluon fields which are often 
neglected due to their highly non-trivial structure. 
Attempts to fix the density dependence of higher order contributions in the 
operator product expansion have e.g. been performed in~\cite{Thomas:2007gx}. 

Here we restrict the present discussion to the scalar 
field $\Sigs$ which follows automatically 
identifying the correlation function with the in-medium nucleon propagator of 
a dressed quasi-particle~\cite{cohen91,drukarev91}
\beqa
\Sigs &=& - \frac{8\pi^2}{\Lambda^{2}_B} 
[\langle{\bar q}q \rangle_{\rho_B}  -\langle {\bar q}q \rangle_0 ]  
= - \frac{8\pi^2}{\Lambda^{2}_B} \frac{\sigma_N}{m_u+m_d} \rho_S .
\label{sum1}
\eeqa
The expression on the right is usually obtained using GOR~(\ref{GOR}) and
the model independent term for the scalar condensate in matter 
which depends linearly on the nucleon density $\rho_B$ in Eq.~(\ref{HF2}).
Consequently the expression is of leading order in density where 
$\rho_S $ is the scalar nucleon density. 
The Borel mass scale $\Lambda_B \simeq 4\pi f_\pi  \simeq 1$ GeV 
is the generic low energy scale of QCD which  separates the 
non-perturbative from the perturbative regime. It coincides with the 
chiral symmetry breaking scale $\Lambda_\chi$ of ChPT. 
Applying Ioffe's formula~\cite{ioffe81} for the nucleon mass 
$M \simeq -\frac{8\pi^2}{\Lambda^{2}_B} \langle {\bar q}q \rangle$ and
the GOR relation one finally obtains ~\cite{finelli}  
\beqa 
\Sigs (\rho) &=& - \frac{\sigma_N M}{m_\pi^2 f_\pi^2} \rho_B~~, 
\label{sum2}
\eeqa 
where the difference between the scalar and vector density, 
$\rho_S$ and $\rho_B$,  can be neglected at low densities $k_F^2 \ll M^2$. 
For the ratio $M^*/M$ follows then
\beqa 
\frac{M^*}{M}=\frac{M+\Sigs}{M}=\frac{\sigma_N}{m_\pi^2 f_\pi^2} \rho_B.
\label{MstarM}
\eeqa 
Naturally, Eq.~(\ref{MstarM}) represents the model independent leading 
order prediction for 
$\langle \bar q q\rangle_{\rho_B}/\langle \bar q q\rangle_0$,~i.e., the 
leading order term in Eq.~(\ref{HF2}). 
Equalizing the two quantities,~i.e., scalar condensate and effective nucleon 
mass, is, however, not as straightforward as Eq.~(\ref{MstarM}) would 
suggest~\cite{birse96,plohl06_2}. As already stated in Ref.~\cite{birse96} 
a direct dependence of the nucleon mass $M^*$ on the in-medium condensate 
contradicts chiral power counting. Moreover, the in-medium quark condensate 
contains contributions from low-momentum virtual pions, which do not 
contribute to the properties of the nucleon in matter.

With the present formalism at hand, we are able to perform a consistent 
comparison of the in-medium scalar condensate, derived from 
the Hellmann-Feynman theorem, and the effective 
nucleon mass where both are obtained from the same chiral EFT interaction 
and at the same order. 

In Fig.~\ref{scalarcond} also the ratio $M^*/M$ is shown with the effective 
nucleon mass given by  $M^*=M+\Sigs$.
The scalar field $\Sigs$ is determined from the chiral EFT potential 
at NLO in HF approximation, making use 
of projection techniques on a relativistic operator basis as described in
Section~\ref{selfenergy}. As one can see also from Table~\ref{tab1}, 
at saturation density the effective mass $M^*$ or the
 ratio $M^*/M$, respectively, is reduced to a value of about 0.546
and is decreasing approximately linear up to $2\rho_0$. 
The reduction of the effective mass $M^*$ is  
about $\approx 13 \%$ larger at $\rho_0$ 
than that of the scalar condensate (HF), 
see also Table~\ref{tab1}. 
At $1.5\,\rho_0$ the difference is about $20\pm 2\%$. Thus the 
approximation of Eq.~(\ref{sum2}) does not hold. 
By a naive comparison of the two quantities the in-medium 
condensate may contribute at the utmost by about $\approx 80\% $
 to the change of the the nucleon mass in matter  at
$1.5\,\rho_0$.
  
As already mentioned, the higher order contributions in Eq.~(\ref{HF3}) from 
the 
nucleon interaction are mainly due to OPE and 
TPE,~i.e., low-momentum virtual pions give the main contribution to the 
change of the scalar quark condensate.
The appearance of the large scalar field $\Sigs$ which enters the effective 
nucleon mass $M^*=M+\Sigs$ originates on the other hand from 
NLO contact interactions (to be more precise from the part which is connected 
to the spin-orbit force),~i.e., it is driven by 
short distance physics~\cite{plohl06_2}. Low-momentum 
pion dynamics is negligible concerning the appearance of the large scalar and 
vector fields $\Sigs$ and $\Sigo$ at the considered order (NLO). 
The present investigations confirm thus 
the considerations of Ref.~\cite{birse96} which were based on a 
chiral expansion of the sigma term.

In summary, a direct dependence of the properties of the nucleon mass on the 
in-medium scalar quark condensate as suggested by Eq.~(\ref{MstarM}) can be 
ruled out.

\section{Summary}
We investigated nuclear bulk properties in the chiral limit $m_\pi \rightarrow 0$. 
This concerns both, the EOS as well as scalar and vector self-energy fields 
in matter. The large relativistic self-energy components are obtained by restoring the
symmetries of the Lorentz group of the nuclear interaction. 
The essential ingredient on which the present investigations 
are based is the chiral EFT nuclear force where the implicit and explicit pion 
mass dependence is known analytically up to NLO and allows a well defined 
extrapolation in the pion mass~\cite{epelbaum03}. 

We found that nuclear bulk properties remain fairly stable in the chiral 
limit. This is true for the EOS as well as the magnitude of the scalar 
and vector mean fields. Both quantities are mainly affected by the pion 
mass dependence of the OPE and TPE. In such investigations a 
source of uncertainty remains  due to the not completely constrained LECs
$\bar D_{S,T}$ which appear in the renormalized contact forces. 
Nevertheless a qualitative change concerning 
the properties of the EOS in the chiral limit 
is not likely to happen.

Furthermore we have calculated the density dependence of the chiral order 
parameter or scalar quark condensate in nuclear matter making use of 
the  Hellmann-Feynman theorem which relates the scalar 
quark condensate with the current quark mass 
derivative of the nuclear energy density. As above, the energy density was 
calculated in  Hartree-Fock approximation. However, to be more realistic 
and to include also short range correlations we applied also the 
Brueckner-Hartree-Fock approximation. 
Since the quark mass dependence of the chiral $NN$ 
interaction is known up to NLO this approach is free from any ambiguities 
which arise concerning the analytic and chiral structure of the potential.
The quark mass dependence of the pion-nucleon coupling constant $g_{\pi N}$ has 
thereby  been taken into account and was found to be important. 
Uncertainties due to unknown low-energy constants entering the pion-nucleon
coupling constant $g_{\pi N} $ do not change the results qualitatively.

Higher order corrections from the nucleon kinetic and interaction 
energy become significantly more important above saturation density 
when compared to the model independent leading order prediction for 
the scalar quark condensate. They lead in general to a weaker 
reduction of the in-medium quark condensate and do not indicate a 
complete restoration of chiral 
symmetry in the density range where hadronic models are reliable.
Since Hartree-Fock and Brueckner-Hartree-Fock provide quantitatively 
comparable results one can conclude that short-range correlations and 
quenching effects, both present in Brueckner theory, have only minor 
implications for the density dependence of the quark condensate.
The substantial contributions from nucleon interactions are due to 
low-momentum virtual pions,~i.e., OPE and TPE. 
Short-distance physics in terms of contact terms and short-range correlations 
have no important impact on the in-medium properties of the quark condensate.

This present formalism allows also to perform a consistent comparison of the 
in-medium scalar condensate, derived directly from the 
Hellmann-Feynman theorem, and the effective nucleon mass $M^*=M+\Sigs$ where 
the scalar self-energy $\Sigs$ enters. For the first time both quantities were 
derived from the same chiral EFT interaction and at the same order. 
In general the effective nucleon mass calculated in the many-body approach is 
smaller ($\approx 10\%$ at $\rho_0$) then the model independent 
leading order prediction 
which is used in the QCD sum rule approach.

Moreover, in the present investigations it turns out, that the reduction 
of the two quantities, namely the in-medium condensate and the 
in-medium nucleon mass, are of 
different physical origin. While the latter is dominantly 
generated by short distance physics in terms of  NLO contact 
interactions~\cite{plohl06_2} virtual low-momentum pions provide the 
essential contributions responsible for the change of the 
in-medium scalar quark condensate.

\acknowledgments
We thank R. Machleidt for providing the Idaho N3LO program 
package and H. Muether for helpful conversation and moreover providing his 
Brueckner-Hartree-Fock code.
This work was supported by the DFG under contract GRK683 (European Graduate 
School T\"ubingen-Basel-Graz). 

%%%%%%%%%%%%%%%%%%%%%%%%%%%%%%%%%%%%%%%%%%%%%%%%%%%%%%%%%%%%%%%%%%%%%%%%

\newpage
%%%%%%%%%%%%%%%%%%%%%%%%%%%%%%%%%%%%%%%%%%%%%%%%%%%%%%%%%%%%%%%%%%%%%%%%%


\begin{thebibliography}{99}

\bibitem{cohen91}
 T.D. Cohen, R.J. Furnstahl, D.K. Griegel, 
Phys. Rev. Lett. 67 (1991) 961; Phys. Rev. C 45 (1992) 1881.

\bibitem{weise90}
S. Klimt, M. Lutz, W. Weise, Phys. Lett. B 249 (1990) 386.

\bibitem{tuebingenC}
T. Maruyama, K. Tsushima and A. Faessler, 
Nucl. Phys. A 535 (1991) 497; 
K. Tsushima, T. Maruyama and A. Faessler, 
Nucl. Phys. A 537 (1992) 303.

\bibitem{Chanfray05}
  G.~Chanfray and M.~Ericson,
 %``QCD Susceptibilities And Nuclear-Matter Saturation In A Relativistic Chiral
  %Theory,''
  Eur.\ Phys.\ J.\ A 25 (2005) 151.


\bibitem{chanfray06}
G. Chanfray, M. Ericson, nucl-th/0611042

\bibitem{Saito05}
  K.~Saito, K.~Tsushima and A.~W.~Thomas,
  %``Nucleon and hadron structure changes in the nuclear medium and impact  on
  %observables,''
  Prog.\ Part.\ Nucl.\ Phys.\ 58 (2007) 1. 

\bibitem{Ratti05}
  C.~Ratti, M.~A.~Thaler and W.~Weise,
  %``Phases of QCD: Lattice thermodynamics and a field theoretical model,''
  Phys.\ Rev.\ D 73 (2006) 014019.

\bibitem{brockmann96}
R. Brockmann, W. Weise, Phys. Lett. B 367 (1996) 40. 

\bibitem{Li94}
G.Q.~Li,C.M.~Ko, Phys. Lett. B 338 (1994) 118.

\bibitem{sw86} 
  B.D. Serot, J.D. Walecka,  
  Adv. Nucl. Phys. 16 (1986) 1.
 
\bibitem{terhaar87a}
B. ter Haar and R. Malfliet,
Phys. Rep. 149 (1987) 207.

\bibitem{bm90}
R. Brockmann, R. Machleidt, Phys. Rev. C 42 (1990) 1965. 

\bibitem{gross99}
T. Gross-Boelting, C. Fuchs, and Amand Faessler,
Nucl. Phys. A 648 (1999) 105.

\bibitem{bonn}
R. Machleidt, K. Holinde, Ch. Elster, Phys. Rep. 149 (1987) 1.

\bibitem{dalen04}
E. van Dalen, C. Fuchs, A. Faessler, Nucl. Phys. A 744 (2004) 227; 
Phys. Rev. Lett. 95 (2005) 022302; 
Phys. Rev. C 72 (2005) 065803.

 
\bibitem{klaehn06}
  T.~Klahn {\it et al.},
  %``Constraints on the high-density nuclear equation of state from the
  %phenomenology of compact stars and heavy-ion collisions,''
  Phys.\ Rev.\ C 74 (2006) 035802. 

\bibitem{entem02} 
D.R. Entem, R. Machleidt, Phys. Rev. C 66 (2002) 014002; 
Phys. Rev. C 68 (2003) 041001.

\bibitem{entem03}
D.R. Entem, R. Machleidt, Phys. Rev. C 68 (2003) 041001.

\bibitem{epelbaum05}
E. Epelbaum, W. Gl\"ockle, Ulf-G. Meissner, 
Nucl. Phys. A 747 (2005) 362. 

\bibitem{epelbaum03}
E. Epelbaum, Ulf-G. Meissner, W. Gloeckle, 
Nucl. Phys A 714 (2003) 535. 

\bibitem{Beane:2002xf}
  S.~R.~Beane and M.~J.~Savage,
  Nucl.\ Phys.\ A 717 (2003) 91. 


\bibitem{epelbaum03b}
E. Epelbaum, W. Gl\"ockle, Ulf-G. Meissner, 
Eur. Phys. J. A 18 (2003) 499. 

\bibitem{drukarev91}
E.G. Drukarev, E.M. Levin, Prog. Part. Nucl. Phys. 27 (1991) 77.

\bibitem{rmf} 
P. Ring, Prog. Part. Nucl. Phys. 73 (1996) 193.

\bibitem{plohl06}
O. Plohl, C. Fuchs, E. van Dalen, Phys. Rev. C 73 (2006) 014003.

\bibitem{plohl06_2}
O. Plohl, C. Fuchs, Phys. Rev. C  74 (2006) 034325.

\bibitem{birse96}
M. Birse, Phys. Rev. C 53  (1996) R2048.

\bibitem{tjon85}
J.A. Tjon, S.J. Wallace,
Phys. Rev. C 32  (1985) 267; Phys. Rev. C 32 (1985) 1667.

\bibitem{horowitz87}
C.J. Horowitz, B.D. Serot,
Nucl. Phys. A 464  (1987) 613.


\bibitem{gross}
F. Gross, J.W. van Orden, K. Holinde,  
Phys. Rev. C 45  (1992) 2094.

\bibitem{Gross:2007be}
  F.~Gross and A.~Stadler,
  %``High-precision covariant one-boson-exchange potentials for np scattering
  %below 350 MeV,''
  arXiv:0704.1229 [nucl-th].

\bibitem{Lehnhart:2004vi}
  B.~C.~Lehnhart, J.~Gegelia and S.~Scherer,
  %``Baryon masses and nucleon sigma terms in manifestly Lorentz-invariant
  %baryon chiral perturbation theory,''
  J.\ Phys.\ G  31 (2005) 89;
  [arXiv:hep-ph/0412092].

\bibitem{machleidt89}
R. Machleidt, Adv. Nucl. Phys. 19  (1989) 189.

\bibitem{weise04}
M. Procura, T. R. Hemmert, W. Weise Phys. Rev. D  69  (2004) 034505.

\bibitem{brown96}
G.E. Brown and M. Rho, Phys. Rep. 269 (1996) 333.

\bibitem{Fuc95}
 C. Fuchs, H. Lenske, and H. H. Wolter,
 Phys. Rev. C 52  (1995) 3043.

\bibitem{Hof01}
F. Hofmann, C. M. Keil, and H. Lenske,
 Phys. Rev. C  64  (2001) 034314. 

\bibitem{dalen06}
  E.~N.~E.~van Dalen, C.~Fuchs and A.~Faessler,
  %``Dirac-Brueckner-Hartree-Fock calculations for isospin asymmetric nuclear
  %matter based on improved approximation schemes,''
  Eur.\ Phys.\ J.\  A  31 (2007) 29.

\bibitem{Typ99}
 S. Typel and H. H. Wolter,
 Nucl. Phys. A 656  (1999) 331; 
 T. Nik$\check{\rm s}$i\'{c}, D. Vretenar, P. Finelli, and P. Ring,
 Phys. Rev. C  66  (2002) 024306.

\bibitem{Nik02}
T. Nik$\check{\textrm{s}}$i$\acute{\textrm{c}}$, D. Vretenar,  
P. Ring,  Phys. Rev. C  66  (2002) 064302. 
 

\bibitem{bulgac56}
A. Bulgac, G. A. Miller, M. Strikman, Phys. Rev. C  56  (1997) 3307.

\bibitem{gasser82}
J.~Gasser and H.~Leutwyler, Phys. Rep.  87  (1982) 77.

\bibitem{gasser91}
J.~Gasser, H.~Leutwyler and M.~E.~Sainio, Phys. Lett. B  253  (1991)252. 

\bibitem{fuchs06} 
C. Fuchs, Prog. Part. Nucl. Phys. 56 (2006) 1.


\bibitem{kuckei02}
J. Kuckei, F. Montani, H. M\"uther, A. Sedrakian, 
Nucl. Phys.  A 723  (2003) 32.

\bibitem{bogner05}
S.K. Bogner, A. Schwenk, R.J. Furnstahl, A. Nogga, 
Nucl. Phys.   A 763  (2005) 59.

\bibitem{said}
SAID on-line program, R.A Arndt, R.L. Workman et al. see website
http://gwdac.phys.gwu.edu/.

\bibitem{matsinos}
E. Matsinos, hep-ph/9807395

\bibitem{koch86}
R. Koch, Nucl. Phys. A 448   (1986) 707.

\bibitem{Thomas:2007gx}
R.~Thomas, T.~Hilger and B.~K\"ampfer,
%``Four-Quark Condensates in Nucleon QCD Sum Rules,''
arXiv:0704.3004 [hep-ph]; 
R. Thomas, S. Zschocke, B. K\"ampfer, 
 Phys. Rev. Lett. 95   (2005) 232301.

\bibitem{ioffe81}
B.L. Ioffe, Nucl. Phys.  B 188   (1981) 317.

\bibitem{finelli} 
P. Finelli, N. Kaiser, D. Vretenar, W. Weise,  
Nucl. Phys.  A 735  (2004) 449. 

\end{thebibliography}
\end{document}